\documentclass%
[english,pra,superscriptaddress,showpacs,a4paper,twocolumn]{revtex4}
\pdfoutput=1
\usepackage[T1]{fontenc}
\usepackage[latin1]{inputenc}
\usepackage{amsmath}
\usepackage{graphicx}
\usepackage{amssymb}
\usepackage{color}
\usepackage{dsfont}
\usepackage{times} \normalfont
\usepackage{babel}
\usepackage{ulem}

\begin{document}
\normalem

\title{
  Turbo charging time-dependent density-functional theory with Lanczos
  chains 
}

\author{
  Dario Rocca
}
\altaffiliation[Present address:]{
  Department of Chemistry, University of California at Davis, Davis,
  California 95616, USA
}
\affiliation{
  SISSA -- Scuola Internazionale Superiore di Studi Avanzati, Via Beirut
  2-4, I-34014 Trieste, Italy
}
\affiliation{
  CNR-INFM DEMOCRITOS National Simulation Center, Trieste, Italy 
}

\author{
  Ralph Gebauer
}
\affiliation{
  ICTP -- The Abdus Salam International Centre for Theoretical Physics,
  Strada Costiera 11, I-34014 Trieste, Italy
}
\affiliation{
  CNR-INFM DEMOCRITOS National Simulation Center, Trieste, Italy 
}

\author{
  Yousef Saad
}
\affiliation{
  Department of Computer Science and Engineering, University of Minnesota,
  and Minnesota Supercomputing Institute, Minneapolis, Minnesota 55455
}

\author{
  Stefano Baroni
}
\affiliation{
  SISSA -- Scuola Internazionale Superiore di Studi Avanzati, Via Beirut
  2-4, I-34014 Trieste, Italy
}
\affiliation{
  CNR-INFM DEMOCRITOS National Simulation Center, Trieste, Italy 
}

\date{\today}

\begin{abstract}
  We introduce a new implementation of time-dependent density-functional
  theory which allows the \emph{entire} spectrum of a molecule or extended
  system to be computed with a numerical effort comparable to that of
  a \emph{single} standard ground-state calculation. This method is
  particularly well suited for large systems and/or large basis sets,
  such as plane waves or real-space grids. By using a super-operator
  formulation of linearized time-dependent density-functional theory,
  we first represent the dynamical polarizability of an interacting-electron
  system as an off-diagonal matrix element of the resolvent of the Liouvillian
  super-operator. One-electron operators and density matrices are treated
  using a representation borrowed from time-independent density-functional
  perturbation theory, which permits to avoid the calculation of unoccupied
  Kohn-Sham orbitals. The resolvent of the Liouvillian is evaluated
  through a newly developed algorithm based on the non-symmetric Lanczos
  method. Each step of the Lanczos recursion essentially requires twice
  as many operations as a single step of the iterative diagonalization
  of the unperturbed Kohn-Sham Hamiltonian. Suitable extrapolation of
  the Lanczos coefficients allows for a dramatic reduction of the number
  of Lanczos steps necessary to obtain well converged spectra, bringing
  such number down to hundreds (or a few thousands, at worst) in typical
  plane-wave pseudopotential applications. The resulting numerical workload
  is only a few times larger than that needed by a ground-state Kohn-Sham
  calculation for a same system. Our method is demonstrated with the
  calculation of the spectra of benzene, C$_{60}$ fullerene, and of
  chlorofyll a.
\end{abstract}


\maketitle

\section{
  Introduction
  \label{sec:Introduction}
}

Time-dependent density-functional theory (TDDFT) \citep{RG:84} stands
as a promising alternative to cumbersome many-body approaches to the
calculation of the electronic excitation spectra of molecular and
condensed-matter systems \citep{Onida:02}. According to a theorem
established by Runge and Gross in the mid eighties \citep{RG:84}, for
any given initial ($t=0$) state of an interacting-electron system, the
external, time-dependent, potential acting on it is uniquely
determined by the time evolution of the one-electron density,
$n(\mathbf{r},t)$, for $t>0$. Using this theorem, one can formally
establish a time-dependent Kohn-Sham (KS) equation from which various
one-particle properties of the system can be obtained as functions of
time. Unfortunately, if little is known about the exchange-correlation
(XC) potential in ordinary density-functional theory (DFT)
\citep{HK:64,KS:65}, even less is known about it in the time-dependent
case. Most of the existing applications of TDDFT are based on the
so-called \emph{adiabatic local density} or \emph{adiabatic
  generalized gradient} approximations (generically denoted in the
following by the acronym ADFT) \citep{Bauernschmitt-Ahlrichs:96},
which amount to assuming the same functional dependence of the XC
potential upon density as in the static case. Despite the crudeness of
these approximations, optical spectra calculated from them are in some
cases almost as accurate as those obtained from more computationally
demanding many-body approaches \citep{Onida:02}. TDDFT is in principle
an exact theory. Progress in understanding and characterizing the XC
functional will substantially increase the predictive power of TDDFT,
while (hopefully) keeping its computational requirements at a
significantly lower level than that of methods based on many-body
perturbation theory.

Linearization of TDDFT with respect to the strength of some external
perturbation to an otherwise time-independent system leads to a
non-Hermitean eigenvalue problem whose eigenvalues are excitation
energies, and whose eigenvectors are related to oscillator strengths
\citep{Casida:95}.  Not surprisingly, this eigenvalue problem has the
same structure that arises in the time-dependent Hartree-Fock theory
\citep{thouless-tdhf-1960,McLachlan:64}, and the dimension of the
resulting matrix (the \emph{Liouvillian}) is twice the product of the
number of occupied (\emph{valence}) states, $N_{v}$, with the number
of unoccupied (\emph{conduction}) states, $N_{c}$. The calculation of
the Liouvillian is by itself a hard task that is often tackled
directly in terms of the unperturbed KS eigen-pairs. This approach
requires the calculation of the full spectrum of the unperturbed KS
Hamiltonian, a step that one may want to avoid when very large basis
sets are used. The diagonalization of the resulting matrix can be
accomplished using iterative techniques \citep{Olsen:1988lr,%
  Stratmann-Scuseria:1998}, often, but not necessarily, in conjunction
with the Tamm-Dancoff approximation \citep{Tamm:45,Dancoff:50,%
  Hirata:99}, which amounts to enforcing Hermiticity by neglecting the
anti-Hermitean component of the Liouvillian. The use of iterative
diagonalization techniques does not necessarily entail the explicit
construction of the matrix to be diagonalized, but just the
availability of a \emph{black-box} routine that performs the product
of the matrix with a test vector (``$H\psi$ products''). An efficient
way to calculate such a product without explicitly calculating the
Liouvillian can be achieved using a representation of the perturbed
density matrix and of the Liouvillian super-operator borrowed from
time-independent density-functional perturbation theory (DFPT)
\citep{BGT:87,Gonze:1995lr,Baroni-RMP:01,furche:tddft-note,hutter-tddft}.
Many applications of TDDFT to atoms, molecules, and clusters have been
performed within such a framework, see for example
Refs. \citep{Bauernschmitt-Ahlrichs:96,Chel-al-TDDFT,furche-rapaport:2005}.
This approach is most likely to be optimal when a small number of
excited states is required. In large systems, however, the number of
quantum states in any given energy range grows with the system
size. The number of pseudo-discrete states in the continuum also grows
with the basis-set size even in a small system, thus making the
calculation of individual eigen-pairs of the Liouvillian more
difficult and not as meaningful. This problem is sometimes addressed
by directly calculating the relevant response function(s), rather than
individual excitation eigen-pairs \citep{Petersilka-Gross:96,Onida:02,%
  Olsen:1988lr,furche:tddft-note}. The price paid in this case is the
calculation and further manipulation (inversion, multiplication) of
large matrices for any individual frequency, a task which may again be
impractical for large systems/basis sets, particularly when an
extended portion of a richly structured spectrum is sought after. For
these reasons, a method to model the absorption spectrum
directly---without calculating individual excited states and not
requiring the calculation, manipulation, and eventual disposal of
large matrices---would be highly desirable.

Such an alternative approach to TDDFT, which avoids diagonalization
altogether, was proposed by Yabana and Bertsch
\citep{Yabana-Bertsch:96}. In this method, the TDDFT KS equations are
solved in the time domain and susceptibilities are obtained by Fourier
analyzing the response of the system to appropriate perturbations in
the linear regime. This scheme has the same computational complexity
as standard time-independent (ground-state) iterative methods in
DFT. For this reason, real-time methods have recently gained
popularity in conjunction with the use of pseudopotentials (PP's) and
real-space grids 
\citep{Marques:2003}, and a similar success should be expected using
plane-wave (PW) basis sets \citep{benzene-Qian,ralph-uspp}. The main
limitation in this case is that, because of stability requirements,
the time step needed for the integration of the TDDFT KS equations is
very small (of the order of $10^{-3}$ fs in typical pseudopotential
applications) and decreasing as the inverse of the PW kinetic-energy
cutoff (or as the square of the real-space grid step)
\citep{ralph-uspp}. The resulting number of steps necessary to obtain
a meaningful time evolution of the TDDFT KS equations may be
exceedingly large.

In a recent letter a new method was proposed \citep{SB-Walker:06} to
calculate optical spectra in the frequency domain---thus avoiding any
explicit integration of the TDDFT KS equations---which does not
require any diagonalization (of either the unperturbed KS Hamiltonian,
or the TDDFT Liouvillian), nor any time-consuming matrix operations.
Most important, the full spectrum is obtained at once without
repeating time-consuming operations for different frequencies. In this
method, which is particularly well suited for large systems and PW, or
real-space grid, basis sets, a generalized susceptibility is
represented by a matrix element of the resolvent of the Liouvillian
super-operator, defined in some appropriate operator space. This
matrix element is then evaluated using a Lanczos recursion
technique. Each link of the Lanczos chain---that is calculated once
for all frequencies---requires a number of floating-point operations
which is only twice as large as that needed by a single step of the
iterative calculation of a \emph{static} polarizability within
time-independent DFPT \citep{BGT:87,Baroni-RMP:01,Gonze:1995lr}. This
number is in turn the same as that needed in a single step of the
iterative diagonalization of a ground-state KS Hamiltonian, or a
single step of Car-Parrinello molecular dynamics.

The purpose of the present paper is to provide an extended and
detailed presentation of the method of Ref. \citep{SB-Walker:06} and
to introduce a few methodological improvements, including a new and
more efficient approach to the calculation of off-diagonal elements of
the resolvent of a non Hermitean operator, and an extrapolation
technique that allows one to substantially reduce the number of
Lanczos recursion steps needed to calculate well converged optical
spectra. The paper is organized as follows. In
Sec. \ref{sec:linearized-Liouville} we introduce the linearized
Liouville equation of TDDFT, including the derivation of an expression
for generalized susceptibilities in terms of the resolvent of the
Liovillian super-operator, the DFPT representation of response
operators and of the Liouvillian super-operator, and the extension of
the formalism to ultrasoft PP's \citep{USPP-VdB:90}; in
Sec. \ref{sec:Lanczos} we describe our new Lanczos algorithm for
calculating selected matrix elements of the resolvent of the
Liouvillian super-operator; in Sec. \ref{sec:benchmark-enhance} we
present a benchmark of the numerical performance of the new method,
and we introduce an extrapolation technique that allows for an
impressive enhancement of it; Sec. \ref{sec:applications} contains
applications of the new methodology to the spectra of
$\mathrm{C}_{60}$ fullerene and to chlorofyll a;
Sec. \ref{sec:conclusions} finally contains our conclusions.

\section{
  Linearized time-dependent density-functional theory 
  \label{sec:linearized-Liouville}
}

The time-dependent KS equations of TDDFT read \citep{RG:84}: 
\begin{equation}
  i\frac{\partial\varphi_{v}({\bf \textbf{r}},t)}{\partial
    t}=\hat{H}_{KS}(t)\varphi_{v}({\bf
    \textbf{r}},t),\label{eq:TD-KSeqs}
\end{equation} 
where 
\begin{equation}
  \hat{H}_{KS}(t)=-\frac{1}{2}\frac{\partial^{2}}{\partial{\bf
      \textbf{r}}^{2}}+v_{ext}({\bf
    \textbf{r}},t)+v_{HXC}(\mathbf{r},t) \label{eq:TD-H_KS}
\end{equation} 
is a time-dependent KS Hamiltonian, $v_{ext}({\bf \textbf{r}},t)$ and
$v_{HXC}({\bf \textbf{r}},t)$ being the time-dependent external and
Hartree plus XC potentials, respectively. In the above equation, as
well as in the following, quantum-mechanical operators are denoted by
a hat, {}``$\hat{\hbox to10pt{\hfill}}$'', and Hartree atomic units
($\hbar=m=e=1$) are used. When no confusion can arise, local
operators, such as one-electron potentials, $\hat{V}$, will be
indicated by the diagonal of their real-space representation, $v({\bf
  \textbf{r}})$, as in Eq.~(\ref{eq:TD-H_KS}).

Let us now assume that the external potential is split into a
time-independent part, $v_{ext}^{\circ}({\bf \textbf{r}})$, plus a
time-dependent perturbation, $v'_{ext}({\bf \textbf{r}},t)$, and let
us assume that the $\varphi$'s satisfy the initial conditions: 
% \YS{Do you really mean boundary conditions instead of ``initial
% conditions'' (condition for t=0)?} 
\begin{equation}
  \varphi_{v}(\mathbf{r},0)=\varphi_{v}^{\circ}({\bf
    \textbf{r}}),\label{eq:BC}
\end{equation} 
where $\varphi_{v}^{\circ}$ are ground-state eigenfunctions of the
unperturbed KS Hamiltonian, $\hat{H}_{KS}^{\circ}$: 
\begin{equation}
  \hat{H}_{KS}^{\circ}\varphi_{v}^{\circ}( \mathbf{r}) = \varepsilon_{v}
  \varphi_{v}^{\circ}(\mathbf{r}).\label{eq:KS0}
\end{equation}
To first order in the perturbation, the time-dependent KS equations
can be cast into the form: 
\begin{multline}
  i\frac{\partial\varphi'_{v}({\bf \textbf{r}},t)} {\partial t}= 
  \left 
    (
    \hat{H}_{KS}^{\circ}-\varepsilon_{v}^{\circ}
  \right
  )\varphi'_{v}({\bf \textbf{r}},t)+\\
  \bigl(v'_{ext}({\bf \textbf{r}},t)+v'_{HXC}({\bf
    \textbf{r}},t)\bigr)\varphi_{v}^{\circ}({\bf
    \textbf{r}}),\label{eq:KS'}
\end{multline} 
where 
\begin{equation}
  \varphi_{v}'(\mathbf{r},t)=\mathrm{e}^{i\varepsilon_{v}t}
  \varphi_{v}(\mathbf{r},t) - \varphi_{v}^{\circ}(\mathbf{r})
  \label{eq:phi'} 
\end{equation}
are the \emph{orbital response functions,} which can be chosen to be
orthogonal to all of the unperturbed occupied orbitals,
$\{\varphi_{v}^{\circ}\}$.

Eq. (\ref{eq:TD-KSeqs}) can be equivalently expressed in terms of
a quantum Liouville equation: 
\begin{equation}
  i\frac{d\hat{\rho}(t)}{dt}=\left[\hat{H}_{KS}(t),\hat{\rho}(t)\right],
  \label{eq:TD-liouville}
\end{equation} 
where $\hat{\rho}(t)$ is the reduced one-electron KS density matrix
whose kernel reads: 
\begin{equation}
  \rho({\bf \textbf{r}},{\bf \textbf{r}}';t) = \sum_{v=1}^{N_{v}}
  \varphi_{v}({\bf
    \textbf{r}},t)\varphi_{v}^{*}(\mathbf{r}',t),
  \label{eq:TD-rho}
\end{equation} 
and the square brackets indicate a commutator. Linearization of
Eq.~(\ref{eq:TD-liouville}) with respect to the external perturbation
leads to: 
\begin{multline}
  i\frac{d\hat{\rho}'(t)}{dt}=
  \left
    [\hat{H}_{KS}^{\circ},\hat{\rho}'(t)
  \right ]+
  \left[
    \hat{V}'_{HXC}(t),\hat{\rho}^{\circ}
  \right] \\
  \left[
    \hat{V}'_{ext}(t),\hat{\rho}^{\circ}
  \right] 
  + \mathcal{O}\left (v'^2\right ) ,\label{eq:delta-liouville}
\end{multline}
where $\hat\rho^\circ$ is the unperturbed density matrix,
$\hat\rho'(t)=\hat\rho(t)-\hat\rho^\circ$, $\hat{V}'_{ext}$ is the
perturbing external potential, and 
$\hat{V}'_{HXC}$ is the variation of the Hartree plus XC potential
linearly induced by $n'(\mathbf{r},t)=\rho'(\mathbf{r},\mathbf{r};t)$:
\begin{multline}
  v'_{HXC}(\mathbf{r},t) = \\ \int 
  \left (
    \frac{1}{|\mathbf{r}-\mathbf{r}'|} \delta(t-t') + \frac{\delta
      v_{XC}(\mathbf{r},t)}{\delta n(\mathbf{r}',t')}
  \right ) 
  n'(\mathbf{r}',t')d\mathbf{r}'dt'. \label{eq:v'-nonadiab}
\end{multline} 
In the ADFT, the functional derivative of the XC potential is assumed
to be local in time,
$ \frac{\delta v_{XC}(\mathbf{r},t)}{\delta n(\mathbf{r}',t')}=
\kappa_{XC}(\mathbf{r},\mathbf{r}')\delta(t-t')$, where
$\kappa_{XC}(\mathbf{r},\mathbf{r}')$ is the functional derivative of
the ground-state XC potential, calculated at the ground-state charge
density, $n^\circ(\mathbf{r})$: $\kappa_{XC}(\mathbf{r},
\mathbf{r}')=\left . \frac{\delta v_{XC} (\mathbf{r})}{\delta
    n(\mathbf{r}')}\right |_{n(\mathbf{r})= n^\circ(\mathbf{r})} $. In
this approximation the perturbation to the XC potential,
Eq. \eqref{eq:v'-nonadiab}, reads therefore:
\begin{equation}
  v'_{HXC}(\mathbf{r},t)=\int \kappa(\mathbf{r},\mathbf{r}')
  n'(\mathbf{r}',t)   d\mathbf{r}',
  \label{eq:v'-adiab}
\end{equation}
where $\kappa(\mathbf{r},\mathbf{r}') = \frac{1}{|\mathbf{r} -
  \mathbf{r}'|} + \kappa_{XC}(\mathbf{r},\mathbf{r}')$. By inserting 
Eq. \eqref{eq:v'-adiab} into Eq. \eqref{eq:delta-liouville}, the
linearized Liouville equation is cast into the form:
\begin{equation}
  i\frac{d\hat{\rho'}(t)}{dt}=\mathcal{L}\cdot\hat{\rho}'(t) +
  \left[\hat{V}'_{ext}(t),\hat{\rho}^{\circ}\right],
  \label{eq:linear-liouville}
\end{equation} 
where the action of the \emph{Liouvillian super-operator},
$\mathcal{L}$, onto $\hat{\rho}'$, $\mathcal{L}\cdot\hat{\rho}',$ is
defined as:
\begin{equation}
  \mathcal{L}\cdot\hat{\rho}' \doteq
  \left[\hat{H}_{KS}^{\circ},\hat{\rho}'\right]
  +\left[\hat{V}'_{HXC}[\hat{\rho}'],\hat{\rho}^{\circ}\right],
  \label{eq:linear-liouvillian}
\end{equation}
and $\hat{V}'_{HXC}[\hat{\rho}']$ is the linear operator functional of 
$\hat{\rho}'$ whose (diagonal) kernel is given by
Eq. \eqref{eq:v'-adiab}.  By Fourier analysing
Eq. \eqref{eq:linear-liouville} we obtain:
\begin{equation}
  (\omega-\mathcal{L})\cdot\tilde{\rho}'(\omega)=
  \left[\tilde{V}'_{ext}(\omega),\hat{\rho}^{\circ}\right],
  \label{eq:super-linear-system}
\end{equation}
where the tilde indicates the Fourier transform and the hat, which
denotes quantum operators, has been suppressed in $\tilde{\rho}'$ and
$\tilde{V}'_{ext}$ in order to keep the notation simple. In the
absence of any external perturbations ($\tilde{V}_{ext}(\omega)=0$),
Eq. (\ref{eq:super-linear-system}) becomes an eigenvalue equation for
$\hat{\rho}'$, whose eigen-pairs describe free oscillations of the
system, \emph{i.e.} excited states \citep{Casida:95}. Eigenvalues
correspond to excitation energies, whereas eigenvectors can be used to
calculate transition oscillator strengths, and/or the response of
system properties to any generic external perturbation.

One is hardly interested in the response of the more general property
of a system to the more general perturbation. When simulating the
results of a specific spectroscopy experiment, one is instead usually
interested in the response of a \emph{specific} observable to a
\emph{specific} perturbation. The expectation value of any
one-electron operator can be expressed as the trace of its product
with the one-electron density matrix. The Fourier transform of the
dipole linearly induced by the perturbing potential, $\hat{V}'_{ext}$,
for example, reads therefore:
\begin{equation}
  \mathbf{d}(\omega)=\mathrm{Tr}
  \left(\hat{\mathbf{r}}\tilde{\rho}'(\omega)\right),
  \label{eq:dipole-1}
\end{equation}
where $\hat{\mathbf{r}}$ is the quantum-mechanical position operator,
and $\hat{\rho}'$ is the solution to
Eq. (\ref{eq:super-linear-system}).  Let us now suppose that the
external perturbation is a homogeneous electric field:
\begin{equation}
  \tilde{v}'_{ext}(\mathbf{r},\omega)=-\mathbf{E}(\omega)
  \cdot\mathbf{r}.
  \label{eq:vE}
\end{equation}
The dipole given by Eq. \eqref{eq:dipole-1} reads therefore:
\begin{equation}
  d_{i}(\omega)=\sum_{j}\alpha_{ij}(\omega)E_{j}(\omega),
  \label{eq:alpha-def}
\end{equation}
where the dynamical polarizability, $\alpha_{ij}(\omega)$, is defined by:
\begin{equation}
  \alpha_{ij}(\omega)=-\mathrm{Tr}\left(\hat{r}_{i}(\omega-\mathcal{L})^{-1}
    \cdot\left[\hat{r}_{j},\hat{\rho}^{\circ}\right]\right).
  \label{eq:alpha-1}
\end{equation}
Traces of products of operators can be seen as scalar products defined
on the linear space of quantum mechanical operators. Let $\hat{A}$ and
$\hat{B}$ be two general one-electron operators. We define their
\emph{scalar product} as:
\begin{equation}
  \bigl\langle\hat{A}|\hat{B}\bigr\rangle\doteq
  \mathrm{Tr}\left(\hat{A}^{\dagger}B\right).
  \label{eq:opsp}
\end{equation}
Eq. \eqref{eq:alpha-1} can thus be formally written as:
\begin{equation} \alpha_{ij}(\omega)=-\left\langle
    \hat{r}_{i}|(\omega-\mathcal{L})^{-1}\cdot\hat{s}_{j}\right\rangle
  ,\label{eq:alpha-2}
\end{equation} 
where 
\begin{equation}
  \hat{s}_{j}=\left[\hat{r}_{j},\hat{\rho}^{\circ}\right]
  \label{eq:commutator}
\end{equation}
is the commutator between the position operator and the unperturbed
one-electron density matrix. The results obtained so far and embodied
in Eq. \eqref{eq:alpha-2} can be summarized by saying that
\emph{within TDDFT the dynamical polarizabilty can be expressed as an
  appropriate off-diagonal matrix element of the resolvent of the
  Liouvillian super-operator}.  A similar conclusion was reached in
Ref. \citep{furche:tddft-note} in the context of a slightly different
formalism. This statement can be extended in a straightforward way to
the dynamic linear response of any observable to any local
one-electron perturbation. It is worth noticing that the operators
that enter the definition of the scalar product in
Eq. \eqref{eq:alpha-2} are orthogonal because $\hat{r}_{i}$ is
Hermitean and $\hat{s}_{j}$ anti-Hermitean (being the commutator of
two Hermitean operators), and the trace of the product of one
Hermitean and one anti-Hermitean operators vanishes.

\subsection{
  Representation of density matrices and other one-electron operators
  \label{sub:TD-DFPT}
}

The calculation of the polarizability using Eqs. (\ref{eq:alpha-1}) or
(\ref{eq:alpha-2}) implies that we should be able to compute
$(\mathcal{L}-\omega)^{-1}\cdot\left[\hat{r}_{j},\hat{\rho}^{\circ}\right]$
in a super-operator linear system. The latter task, in turn, requires
an explicit representation for the density-matrix response,
$\tilde{\rho}'$, for its commutator with the unperturbed Hamiltonian,
for local operators, such as $\hat{r}_{j}$ of $\hat{V}'_{HXC}$, for
their commutators with the unperturbed density matrix, as well as for
the Liouvillian super-operator, or at least for its product with any
relevant operators, $\hat{A}$, such as $\mathcal{L}\cdot\hat{A}$.

A link between the orbital and density-matrix representations of TDDFT
expressed by Eqs. (\ref{eq:KS'}) and (\ref{eq:delta-liouville}) can be
obtained by linearizing the expression (\ref{eq:TD-rho}) for the
time-dependent density matrix:
\begin{equation} \rho'({\bf r},{\bf
    r}';t)=\sum_{v}\bigl[\varphi_{v}^{\circ}({\bf
    r})\varphi'{}_{v}^{*}({\bf r}',t)+\varphi'{}_{v}({\bf
    r},t)\varphi_{v}^{\circ*}({\bf
    r}')\bigr],\label{eq:rho'(t)}
\end{equation} 
whose Fourier transform reads: 
\begin{multline}
  \tilde{\rho}'({\bf r},{\bf r}';\omega)=\\
  \sum_{v}\bigl[\varphi_{v}^{\circ}({\bf
    r})\tilde{\varphi}'{}_{v}^{*}({\bf
    r}',-\omega)+\tilde{\varphi}'{}_{v}({\bf
    r},\omega)\varphi_{v}^{\circ*}({\bf
    r}')\bigr].\label{eq:rho'(omega)}
\end{multline}
Eq. \eqref{eq:rho'(omega)} shows that $\tilde{\rho}(\omega)$ is
univocally determined by the two sets of orbital response functions,
$\mathbf{x}'=\{\varphi'_{v}(\mathbf{r},\omega)\}$ and
$\mathbf{y}'=\{\varphi'{}_{v}^{*}({\bf r},-\omega)\}$.  A set of a
number of orbitals equal to the number of occupied states, such as
$\mathbf{x}'$ or $\mathbf{y}'$, will be nicknamed a \emph{batch} of
orbitals. Notice that $\tilde{\rho}(\omega)$ is \emph{not} Hermitian
because the Fourier transform of a Hermitian, time-dependent, operator
is not Hermitian, unless the original operator is even with respect to
time inversion.  Because of the orthogonality between occupied and
response orbitals ($\langle\varphi_{v}^{\circ}|
\varphi'_{v'}\rangle=0$), Eq. \eqref{eq:rho'(t)} implies that the
matrix elements of $\hat{\rho}'$ between two unperturbed KS orbitals
which are both occupied or both empty vanish
($\rho'_{vv'}=\rho'_{cc'}=0$), as required by the idempotency of
density matrices $(\hat\rho^2=\hat\rho)$ in DFT. As a consequence, in
order to calculate the response of the expectation values of a
Hermitian operator, $\hat{A}$, such as in Eq. \eqref{eq:dipole-1}, one
only needs to know and represent the occupied-empty (\emph{vc}) and
empty-occupied (\emph{cv}) matrix elements of $\hat{A}$, $A_{vc}$ and
$A_{cv}$. In other terms, if we define as
$\hat{P}=\sum_{v}|\varphi_{v}^{\circ} \rangle \langle
\varphi_{v}^{\circ}| \doteq\hat{\rho}^{\circ}$ and
$\hat{Q}\doteq1-\hat{P}$ as the projectors onto the occupied- and
empty-state manifolds, respectively, one has that:
\begin{equation}
  \mathrm{Tr}(\hat{A}\tilde{\rho}'(\omega))=
  \mathrm{Tr}(\hat{A}'\tilde{\rho}'(\omega)),
  \label{eq:TraAAp}
\end{equation}
where $\hat{A}'=\hat{P}\hat{A}\hat{Q}+\hat{Q}\hat{A}\hat{P}$ is the
$vc$-$cv$ component of $\hat{A}$, which can be easily and conveniently
represented in terms of batches of orbitals. To this end, let us
define the orbitals:
\begin{eqnarray}
  a_{v}^{x}(\mathbf{r}) & = &
  \hat{Q}\hat{A}\varphi_{v}^{\circ}(\mathbf{r})=
  \sum_{c}\varphi_{c}^{\circ}(\mathbf{r})A_{cv},\label{eq:batch-1}\\
  a_{v}^{y}(\mathbf{r}) & = & 
  \left(\hat{Q}\hat{A}^{\dagger}\varphi_{v}^{\circ}(\mathbf{r})\right)^{*}=
  \sum_{c}\varphi_{c}^{\circ*}(\mathbf{r})A_{vc}.
  \label{eq:batch-2}
\end{eqnarray}
One has then:
\begin{eqnarray}
  A_{cv} & = & \langle\varphi_{c}^{\circ}|a_{v}^{x}\rangle,
  \label{eq:Acv-def}\\
  A_{vc} & = &
  \langle\varphi_{c}^{\circ*}|a_{v}^{y}\rangle.
  \label{eq:Avc-def}
\end{eqnarray}
If Eqs. (\ref{eq:Acv-def}) and (\ref{eq:Avc-def}) are used to
represent density matrices, then the free oscillations
corresponding to setting $\tilde V'_{ext}=0$ in
Eq. \eqref{eq:super-linear-system} would be described by Casida's
eigenvalue equations \citep{Casida:95}. 

For simplicity and without much loss of generality, from now on we
will assume that the unperturbed system is time-reversal invariant, so
that the unperturbed KS orbitals, $\varphi_{v}^{\circ}$ and
$\varphi_{c}^{\circ}$, can be assumed to be real. The two batches of
orbitals $\mathbf{a}^{x}\doteq\{ a_{v}^{x}(\mathbf{r})\}$ and
$\mathbf{a}^{y}=\{ a_{v}^{y}(\mathbf{r})\}$ will be called the
\emph{batch representation} of the $\hat{A}$ operator, and indicated
with the notation $(\mathbf{a}^{x},\mathbf{a}^{y})$ or $(\{
a_{v}^{x}\},\{ a_{v}^{y}\}).$ Scalar products between operators
(traces of operator products) can be easily expressed in terms of
their batch representations. Let $(\{ b_{v}^{x}\},\{ b_{v}^{y}\})$ be
the batch representation of the operator $\hat{B}$. If either of the
two operators, $\hat{A}$ or $\hat{B}$, has vanishing $vv$ and $cc$
components, one has:
\begin{eqnarray}
  \left<\hat{A}|\hat{B}\right> & = & 
  \mathrm{Tr}\left(\hat{A}^{\dagger}B\right)\nonumber \\
  & = & \sum_{cv}\left(A_{cv}^{*}B_{cv}+A_{vc}^{*}B_{vc}\right)\nonumber \\
  & = & \sum_{v}\left(\langle a_{v}^{x}|b_{v}^{x}\rangle+\langle
    a_{v}^{y}|b_{v}^{y}\rangle\right).
  \label{eq:scalar-batch}
\end{eqnarray}
If $\hat{A}$ is Hermitian, its batch representation satisfies the
relation: $a^{y}(\mathbf{r})=a^{x}(\mathbf{r})^{*}$, whereas
anti-Hermiticity would imply: $a^{y}(\mathbf{r})=-a^{x}(
\mathbf{r})^{*}$. Due to time-reversal invariance and the consequent
reality of the unperturbed KS orbitals, the batch representation of a
real (imaginary) operator is real (imaginary), and the batch
representation of a local operator, $\hat{V}$ (which is Hermitean,
when real, or non Hermitean, when complex), satisfies:
$v_{v}^{y}(\mathbf{r})=v_{v}^{x}(\mathbf{r})$.

In order to solve the super-operator linear system,
Eq. (\ref{eq:super-linear-system}), using the batch representation,
one needs to work out the batch representation of
$\tilde{V}'_{HXC}(\mathbf{r,}\omega)$ as a functional of
$\tilde{\rho}'$, as well as of the various commutators appearing
therein. The charge-density response to an external perturbation
reads:
\begin{eqnarray} n'(\mathbf{r}) & = &
  \sum_{v}\varphi_{v}^{\circ}({\bf
    r})\bigl(\tilde{\varphi}'{}_{v}({\bf
    r},\omega)+\tilde{\varphi}'{}_{v}^{*}({\bf
    r}',-\omega)\bigr)\nonumber \\
  & = & \sum_{v}\varphi_{v}^{\circ}({\bf
    r})\bigl(x'_{v}(\mathbf{r})+y'_{v}(\mathbf{r})\bigr),
  \label{eq:n'-xy}
\end{eqnarray} 
where $(\{ x_{v}'\},\{ y_{v}'\})$ is the batch representation of the
density-matrix response, $\tilde{\rho}'$. The Hartree-plus-XC
potential response is:
\begin{eqnarray}
  v'_{HXC}[\tilde{\rho}'](\mathbf{r}) & = &
  \int\kappa(\mathbf{r},\mathbf{r}')n'(\mathbf{r}')d\mathbf{r}'\nonumber
  \\ 
  & = &
  \sum_{v}\int\kappa(\mathbf{r},\mathbf{r}')\varphi_{v}^{\circ}
  (\mathbf{r}')\nonumber \\ 
  & &
  \quad\quad\quad\quad\times\bigl(x'_{v}(\mathbf{r}')+
  y'_{v}(\mathbf{r}')\bigr)d\mathbf{r}'.
  \label{eq:V'HXC-xy}
\end{eqnarray}
Using Eqs. \eqref{eq:batch-1} and \eqref{eq:batch-2} the batch
representation of the Hartree-plus-XC potential response reads
therefore:
\begin{eqnarray}
  v'{}_{HXC,v}^{x}(\mathbf{r}) & = &
  \hat{Q}\sum_{v'}\int\varphi_{v}^{\circ}(\mathbf{r})
  \kappa(\mathbf{r},\mathbf{r}')\varphi_{v'}^{\circ}(\mathbf{r}')\nonumber
  \\ 
  &  &
  \quad\quad\quad\quad\times\bigl(x'_{v'}(\mathbf{r}')+
  y'_{v'}(\mathbf{r}')\bigr)d\mathbf{r}'\\ 
  & \doteq & \hat{Q}\sum_{v'}\int K_{vv'}(\mathbf{r},\mathbf{r}')\nonumber \\
  &  & \quad\quad\quad\quad\times
  \bigl(x'_{v'}(\mathbf{r}')+y'_{v'}(\mathbf{r}')\bigr)d\mathbf{r}'
  \label{eq:v'hxc-batch-x}\\
  v'{}_{HXC,v}^{y}(\mathbf{r}) & = &
  v'{}_{HXC,v}^{x}(\mathbf{r}),\label{eq:v'hxc-batch-y}\end{eqnarray}
where:
\begin{equation}
  K_{vv'}(\mathbf{r},\mathbf{r}')=
  \kappa(\mathbf{r},\mathbf{r'}) \varphi_{v}^{\circ}
  (\mathbf{r})\varphi_{v'}^{\circ}(\mathbf{r}')
  \label{eq:Kvvp}
\end{equation}
[See Eq. \eqref{eq:v'-adiab}]. Let $(\{ v'{}{}_{v}^{x}\},\{
v'{}{}_{v}^{y}\})$ be the batch representation of a local operator,
$\hat{V}'$. The batch representation of the commutator between
$\hat{V}'$ and the unperturbed density matrix,
$\hat{V}''=[\hat{V}',\hat{\rho}^{\circ}]$, reads:
\begin{eqnarray}
  v''{}_{v}^{x}(\mathbf{r}) & = &
  \hat{Q}[\hat{V}',\hat{\rho}^{\circ}]\varphi_{v}^{\circ}(\mathbf{r}) \nonumber
  \\ 
  & = & v'{}_{v}^{x}(\mathbf{r})\label{eq:v''-x}\\
  v''{}_{v}^{y}(\mathbf{r}) & = &
  -v''{}_{v}^{x}(\mathbf{r}).
  \label{eq:v''-y}
\end{eqnarray} 
The batch
representation of the commutator between the unperturbed Hamiltonian
and the density-matrix response,
$\tilde{\rho}''=[\hat{H}^{\circ},\tilde{\rho}']$,
reads:
\begin{eqnarray}
  x_{v}''(\mathbf{r}) & = &
  \hat{Q}[\hat{H}^{\circ},\tilde{\rho}']\varphi_{v}^{\circ}(\mathbf{r})\nonumber
  \\ 
  & = & (\hat{H}^{\circ}-\varepsilon_{v})x_{v}'(\mathbf{r})\label{eq:x''}\\
  y_{v}''(\mathbf{r}) & = &
  -(\hat{H}^{\circ}-\varepsilon_{v})y_{v}'(\mathbf{r}).
  \label{eq:y''}
\end{eqnarray}
The batch representation of the product of the Liouvillian with the
density-matrix response $(\mathcal{L}\cdot \tilde\rho')$ appearing in
Eq. \eqref{eq:super-linear-system} reads:
\begin{equation}
  \mathcal{L}
  \left(
    \begin{array}{c}
      \mathbf{x}'\\
      \mathbf{y}'
    \end{array}
  \right)
  =
  \left(
    \begin{array}{cc}
      \mathcal{D}+\mathcal{K} & \mathcal{K}\\
      -\mathcal{K} &
      -\mathcal{D}-\mathcal{K}
    \end{array}
  \right)
  \left(
    \begin{array}{c}
      \mathbf{x}'\\
      \mathbf{y}'
    \end{array}
  \right),
  \label{eq:super-L}
\end{equation}
where the action of the $\mathcal{D}$ and $\mathcal{K}$
super-operators on batches of orbitals is defined as: 
\begin{eqnarray}
  \mathcal{D}\{ x_{v}(\mathbf{r})\} & = &
  \{(\hat{H}^{\circ}-\varepsilon_{v})x_{v}(\mathbf{r})\}\label{eq:super-D}\\ 
  \mathcal{K}\{ x_{v}(\mathbf{r})\} & = & \left\{ \hat{Q}\sum_{v'}\int
    K_{vv'}(\mathbf{r},\mathbf{r}')x_{v'}(\mathbf{r}')d\mathbf{r}'\right\}
  .
  \label{eq:super-K}
\end{eqnarray} 
Note that, according to Eqs. \eqref{eq:super-L}, \eqref{eq:super-D},
and \eqref{eq:super-K}, the calculation of the product of the
Liouvillian with a general one-electron operator in the batch
representation only requires operating on a number of one-electron
orbitals equal to the number of occupied KS states (number of
electrons), without the need to calculate any empty states. In
particular, the calculation of Eq.  \eqref{eq:super-K} is best
performed by first calculating the HXC potential generated by the
fictitious charge density
$\bar{n}(\mathbf{r})=\sum_{v}x_{v}(\mathbf{r})\varphi_{v}^{\circ}(\mathbf{r})$,
and then applying it to each unperturbed occupied KS orbital,
$\varphi_{v}^{\circ}(\mathbf{r}).$ The projection of the resulting
orbitals onto the empty-state manifold implied by the multiplication
with $\hat{Q}$ is easily performed using the identity:
$\hat{Q}=1-\sum_{v}|\varphi_{v}^{\circ}\rangle\langle\varphi_{v}^{\circ}|$,
as it is common practice in DFPT.

Following Tsiper \citep{tsiper-2001}, it is convenient to perform
a $45^{\circ}$ rotation in the space of batches and define:
\begin{eqnarray}
  q_{v}(\mathbf{r}) & = &
  \frac{1}{2}\bigl(x_{v}(\mathbf{r})+y_{v}(\mathbf{r})\bigr)
  \label{eq:batch-q}\\ p_{v}(\mathbf{r}) & = &
  \frac{1}{2}\bigl(x_{v}(\mathbf{r})-y_{v}(\mathbf{r})\bigr).
  \label{eq:batch-p}
\end{eqnarray} 
Eqs. \eqref{eq:batch-q} and \eqref{eq:batch-p} define the \emph{standard
batch representation} (SBR) of the density-matrix response. The SBR
of the response charge density is:
\begin{equation}
  n'(\mathbf{r})=2\sum_{v}\varphi_{v}^{\circ}(\mathbf{r})q_{v}(\mathbf{r}).
  \label{eq:n'-SBR}
\end{equation}
The SBR of a general one-electron operator is defined in a similar
way. In particular, the SBR of a real Hermitian operator has zero
$p$ component, wheres the SBR of the commutator of such an operator
with the unperturbed density matrix has zero $q$ component. The
standard batch representation of the TDDFT Liouville equation,
Eq. \eqref{eq:super-linear-system},
reads:
\begin{equation}
  \left(
    \begin{array}{cc}
      \omega & -\mathcal{D}\\
      -\mathcal{D}-2\mathcal{K} & \omega
    \end{array}
  \right)
  \left(
    \begin{array}{c}
      \mathbf{q}'\\
      \mathbf{p}'
    \end{array}
  \right)
  =
  \left(
    \begin{array}{c}
      0\\
      \{\hat{Q}v_{ext}(\mathbf{r})\varphi_{v}^{\circ}(\mathbf{r})\}
    \end{array}
  \right).
  \label{eq:sup-1}
\end{equation}

In conclusion, the batch representation of response density matrices
and of general one-electron operators allows one to avoid the explicit
calculation of unoccupied KS states, as well as of the Liouvillian matrix,
which is mandatory when (very) large one-electron basis sets (such as
PW's or real-space grids) are used to solve the ground-state problem.
This representation is the natural extension to the time-dependent
regime of the practice that has become common since the introduction
of time-independent DFPT\hbox to
0pt{\phantom{\citep{Ochsenfeld:1997lr,hutter-tddft}}\hss}
\citep{Baroni-RMP:01,BGT:87,batch-QC}.

\subsection{
  Ultra-soft pseudopotentials
  \label{sub:uspp}
}

The formalism outlined above applies to all-electron as well as to
pseudopotential calculations performed using norm-conserving
pseudopotentials, which give rise to an ordinary KS ground-state
eigenvalue problem. Ultra-soft pseudo-potentials (USPP's)
\citep{USPP-VdB:90}, instead, give rise to a generalized KS
ground-state eigenvalue problem and the time evolution within TDDFT
has to be modified accordingly \citep{benzene-Qian,ralph-uspp}. The
generalization of the TDDFT formalism to USPP's has been presented in
full detail in Ref. \citep{ralph-uspp}, and here we limit ourselves to
report the main formulas.

In the framework of USPP's, the charge density is written as a sum
$n(\mathbf{r},t)=n^{\text{US}}(\mathbf{r},t)+n^{\text{aug}}(\mathbf{r},t)$.
The delocalized contribution, $n^{\text{US}}$, is represented as the sum
over the squared moduli of the KS orbitals:
$n^{\text{US}}(\mathbf{r},t) =
\sum_{v}\left|\varphi_{v}(\mathbf{r},t)\right|^{2}$. The {\em
  augmentation charge}, $n^{\text{aug}}$, instead, is written in terms
of so-called augmentation functions $Q_{nm}^{I}(\mathbf{r})$:
\begin{equation}
  n^{\text{aug}}(\mathbf{r},t) = \sum_{v}\sum_{n,m,I}
  Q_{nm}^{I}(\mathbf{r}) \langle\varphi_{v}(t)| \beta_{n}^{I}\rangle
  \langle\beta_{m}^{I}|\varphi_{v}(t)\rangle.
  \label{eq:naug}
\end{equation}
The augmentation functions, as well as the functions
$\beta_{n}^{I}(\mathbf{r})\doteq\beta_{n}(\mathbf{r}-\mathbf{R}_{I})$
are localized in the core region of atom $I$. The $\beta$'s each
consist of an angular momentum eigenfunction times a radial function
that vanishes outside the core radius. Typically one or two such
functions are used for each angular momentum channel and atom
type. The indices $n$ and $m$ in Eq.~(\ref{eq:naug}) run over the
total number of such functions for atom $I$. In practice, the
functions $Q_{nm}(\mathbf{r})$ and $\beta_{n}(\mathbf{r})$ are
provided with the pseudopotential for each type of atom.

The advantage of using USPP's over standard norm-conserving
pseudopotentials comes from this separation of the strongly localized
contributions to the charge density from the more delocalized
contributions. The square moduli of the KS orbitals only represent the
latter part of $n(\mathbf{r},t)$, and therefore fewer Fourier
components in the representation of the orbitals are sufficient for a
correct representation of the charge density. As a consequence the
kinetic energy cutoff which determines the size of the basis set can
be chosen much smaller in typical USPP applications than in
corresponding calculations with norm-conserving PP's. As shown in
Ref.~\onlinecite{ralph-uspp}, the smaller basis set not only reduces
the dimensions of the matrices during the computation, but it allows
also for a faster convergence of spectroscopic quantities, when
calculated both with real-time or with spectral Lanczos techniques
(see Sec. \ref{sec:Lanczos}).

The generalized expression for the USPP charge density given above
entails a more complicated structure of the KS eigenvalue problem.
Instead of the standard eigenvalue equation (\ref{eq:KS0}), one now
has 
\begin{equation}
  \hat{H}_{KS}^{\circ}\varphi_{v}^{\circ}(\mathbf{r}) =
  \varepsilon_{v}\hat{S}\varphi_{v}^{\circ}(\mathbf{r}),
  \label{eq:KSUSPP}
\end{equation} 
where the overlap operator $\hat{S}$ is defined as 
\begin{equation}
  \hat{S}=\hat{\mathds{1}} + \sum_{n,m,I}q_{nm}^{I}|
  \beta_{n}^{I}\rangle
  \langle\beta_{m}^{I}|,
  \label{eq:Sop}
\end{equation}
with $q_{nm}^{I}=\int d\mathbf{r}\, Q_{nm}^{I}(\mathbf{r})$ and
$\hat{\mathds{1}}$ the identity operator.
Consequently, the equation for the time-dependent KS orbitals,
Eq.~(\ref{eq:KS'}), also contains the overlap operator in the USPP
formalism: 
\begin{multline}
  i\hat{S}\frac{\partial\varphi'_{v}({\bf \textbf{r}},t)}{\partial t}
  =\left(\hat{H}_{KS}^{\circ}-\hat{S}\varepsilon_{v}^{\circ}\right)
  \varphi'_{v}({\bf \textbf{r}},t)+\\
  \bigl(v'_{ext}({\bf \textbf{r}},t)+v'_{HXC}({\bf
    \textbf{r}},t)\bigr)\varphi_{v}^{\circ}({\bf
    \textbf{r}}).\label{eq:KSUSPP'}
\end{multline}

Using the same derivation as before, but starting from
Eq.~(\ref{eq:KSUSPP'}) instead of Eq.~(\ref{eq:KS'}), we arrive at a
standard batch representation of the TDDFT Liouville equation in the
USPP formalism. It has the same form as Eq.~(\ref{eq:sup-1}) above,
but with the super-operators ${\mathcal{D}}$ and ${\mathcal{K}}$
replaced by:
\begin{align}
  \mathcal{D}^{US}\{ x_{v}(\mathbf{r})\} & =
  \{(\hat{S}^{-1}\hat{H}^{\circ}-\varepsilon_{v})x_{v}(\mathbf{r})\}
  \label{eq:USsuper-D}\\ 
  \mathcal{K}^{US}\{ x_{v}(\mathbf{r})\} & =
  \left\{ \hat{S}^{-1}\hat{Q}\sum_{v'}\int
    K_{vv'}(\mathbf{r},\mathbf{r}')x_{v'}(\mathbf{r}')d\mathbf{r}'\right\}
  , \label{eq:USsuper-K}
\end{align} 
and the right-hand side of Eq.~(\ref{eq:sup-1}) by 
\begin{equation}
  \left(
    \begin{array}{c}
      0\\
      \{\hat{S}^{-1}\hat{Q}v_{ext}(\mathbf{r})\varphi_{v}^{\circ}(\mathbf{r})\}
    \end{array}
  \right).
\end{equation}
In this case the projector onto the empty-state manifold is defined as
\begin{equation}
\hat{Q}=\hat{S}-\sum_{v}\hat{S}| \varphi_{v}^{\circ}\rangle
\langle\varphi_{v}^{\circ}|.
\end{equation}
The inverse overlap operator, $\hat{S}^{-1}$, appearing in these
expressions can be cast in the form 
\begin{equation}
  \hat{S}^{-1}=\hat{\mathds{1}}+\sum_{n,m,I,J}\lambda_{nm}^{IJ}|
  \beta_{n}^{I}\rangle\langle\beta_{m}^{J}|,
  \label{eq:Sm1}
\end{equation}
which is very similar to the $\hat{S}$ operator itself, given in
Eq.~(\ref{eq:Sop}), except fact that $\hat{S}^{-1}$ generally
connects $\beta$-functions localized on different atoms. The numbers
$\lambda_{nm}^{IJ}$ can be obtained from the condition
$\hat{S}\hat{S}^{-1}=\hat{\mathds{1}}$. If the atoms are kept at fixed
positions, as it is the case here, the overlap operator is independent
of time and the $\lambda$'s need to be calculated only once
for all.

\section{
  Generalized susceptibilities from Lanczos recursion chains 
  \label{sec:Lanczos} 
}

According to Eq. \eqref{eq:alpha-2}, the polarizability can be
expressed as an appropriate off-diagonal matrix element of the
resolvent of the non-Hermitian Liouvillian (super-) operator between
two orthogonal vectors. The standard way to calculate such a matrix
element is to solve first a linear system whose right-hand side is the
\emph{ket} of the matrix element, and to calculate then the scalar
product between the solution of this linear system with the \emph{bra}
\citep{Olsen:1988lr,furche:tddft-note}. The main limitation of such an
approach is that 
%% YS change in wording: the solution of linear system entails the
solving linear systems entails the manipulation and storage of a large
amount of data and that a different linear system has to be solved
from scratch for each different value of the frequency. In the case of
a \emph{diagonal} element of a \emph{Hermitian} operator, a very
efficient method, based on the Lanczos factorization algorithm
\citep[p. 185 and {\em ff.}]{saad-book} is known, which allows to
avoid the solution of the linear system altogether
\citep{haydock-72-1,haydock-75,recursion-ssp,pastori-recursion}.
Using such a method (known as the \emph{Lanczos recursion method}) a
diagonal matrix element of the resolvent of a Hermitean operator can
be efficiently and elegantly expressed in terms of a continued
fraction generated by a Lanczos recursion chain starting from the
vector with respect to which one wants to calculate the matrix element
\citep{haydock-72-1,haydock-75,recursion-ssp,pastori-recursion}. The
generalization of the Lanczos recursion method to non-Hermitian
operators is straightforward, based on the Lanczos biorthogonalization
algorithm \citep[p. 503]{GVL-book}. This generalization naturally
applies to the calculation of an off-diagonal matrix element between
vectors that are not orthogonal.  Less evident is how to encompass the
calculation of off-diagonal matrix elements between orthogonal
vectors. In Ref. \citep{SB-Walker:06} such matrix elements were
treated using a block version of the Lanczos
bi-orthogonalization. This approach has the drawback that a different
Lanczos chain has to be calculated for the response of each different
property to a given perturbation (\emph{i.e.} for each different
\emph{bra} in the matrix element corresponding to a same
\emph{ket}). In the following, we generalize the recursion method of
Haydock, Heine, and Kelly \citep{haydock-72-1,haydock-75,%
  recursion-ssp,pastori-recursion}, so as to encompass the case of an
\emph{off-diagonal} element of the resolvent of a \emph{non-Hermitean}
operator without resorting to a block variant of the algorithm and
allowing to deal with the case in which the left and the right vectors
are orthogonal. This will allow us to calculate the full dynamical
response of \emph{any} dynamical property to a given perturbation,
from a {\em single} scalar Lanczos chain.

We want to calculate quantities such as: 
\begin{equation}
  g(\omega)=\left\langle u|(\omega-\mathcal{A})^{-1}v\right\rangle
  ,\label{eq:resolvent-def}
\end{equation}
where $\mathcal{A}$ is a non-Hermitean matrix defined in some linear
space, whose dimension will be here denoted $n$, and $u$ and $v$ are
elements of this linear space, which we suppose to be normalized:
$\parallel u\parallel=\parallel v\parallel=1$, where $\parallel
v\parallel^{2}=\langle v|v\rangle$.  For simplicity, and without loss
of generality in view of applications to time-reversal invariant
quantum-mechanical problems, we will assume that the linear space is
defined over real numbers. Let us define a sequence of
\emph{left} and \emph{right} vectors\emph{, $\{ p_{1},p_{2},\cdots
  p_{k},\cdots\}$} and \emph{$\{ q_{1},q_{2},\cdots q_{k},\cdots\}$,}
from the following procedure, known as the Lanczos
\emph{bi-orthogonalization} algorithm \citep[p. 503]{GVL-book}:
\begin{eqnarray}
  \gamma_{1}q_{0}=\beta_{1}p_{0} & = & 0,\label{eq:Lanczos-0}\\
  q_{1}=p_{1} & = & v,\label{eq:Lanczos-start}\\
  \beta_{j+1}q_{j+1} & = &
  \mathcal{A}^{\phantom{\top}}q_{j}-\alpha_{j}q_{j}-\gamma_{j}q_{j-1},
  \label{eq:Lanczos-rec1}\\ 
  \gamma_{j+1}p_{j+1} & = &
  \mathcal{A}^{\top}p_{j}-\alpha_{j}p_{j}-\beta_{j}p_{j-1},
  \label{eq:Lanczos-rec2}
\end{eqnarray}  
where: 
\begin{eqnarray} \alpha_{j} & = & \langle
  p_{j}|\mathcal{A}q_{j}\rangle,\label{eq:lanczos-alpha}
\end{eqnarray}
and $\beta_{j+1}$ and $\gamma_{j+1}$ are scaling factors for the
vectors $q_{j+1}$ and $p_{j+1}$, respectively, so that they will
satisfy: 
\begin{equation} \langle
  q_{j+1}|p_{j+1}\rangle=1.\label{eq:bi-norm}
\end{equation} 
Thus, from an algorithmic point of view, the right-hand sides of
Eqs. (\ref{eq:Lanczos-rec1}-\ref{eq:Lanczos-rec2}) are evaluated first
%% YS minor change in wording 
with $\alpha_{j}$ obtained from Eq. \eqref{eq:lanczos-alpha}. Then, the two
scalars $\beta_{j+1}$ and $\gamma_{j+1}$ are determined so that
Eq. \eqref{eq:bi-norm} is satisfied. Eq. \eqref{eq:bi-norm} only gives
a condition on the \emph{product} of $\beta_{j+1}$ and
$\gamma_{j+1}$. If we call $\bar{q}$ and $\bar{p}$ the vectors on the
right-hand sides of Eqs. \eqref{eq:Lanczos-rec1},
\eqref{eq:Lanczos-rec2} respectively, this condition is that
$\beta_{j+1}\gamma_{j+1}=\langle\bar{q}|\bar{p}\rangle$.
In practice one typically sets:
\begin{eqnarray}
  \beta_{j+1} & = & \sqrt{|\langle\bar{q}|\bar{p}\rangle|}
  \label{eq:betaj+1}\\
  \gamma_{j+1} & = & \mbox{sign}(\langle\bar{q}|\bar{p}\rangle)
  \times\beta_{j+1}.
  \label{eq:gammaj+1}
\end{eqnarray}  
The set of $q$ and $p$ vectors thus generated are said to be links
of a \emph{Lanczos chain}. 
%% YS - forgot  to say that the q_i's and p_j' are orthogonal!! 
In exact arithmetics, it is known that these two sequences of vectors 
are mutually orthogonal to each other, i.e., 
$\langle q_i | p_j \rangle = \delta_{ij} $, where 
$ \delta_{ij} $ is  the Kronecker symbol. 

The resulting algorithm is described in detail, \emph{e.g.}, in Refs.
\citep{GVL-book,saad-book}. Let us define $Q^{j}$ and $P^{j}$ as the
$(n\times j)$ matrices: 
\begin{eqnarray}
  Q^{j} & = & [q_{1},q_{2},\cdots,q_{j}],\label{eq:Lanczos-Qj}\\
  P^{j} & = &
  [p_{1},p_{2},\cdots,p_{j}],\label{eq:Lanczos-Pj}
\end{eqnarray} 
and let $e_{k}^{m}$ indicate the $k$-th unit vector in a
$m$-dimensional space (when there is no ambiguity on the
dimensionality of the space, the superscript $j$ will be dropped). The
following Lanczos factorization holds in terms of the quantities
calculated from the recursions equations
(\ref{eq:Lanczos-start}-\ref{eq:Lanczos-rec2}): 
\begin{eqnarray}
  \mathcal{A}^{\phantom{\top}}Q^{j} & = &
  Q^{j}T^{j\phantom{\top}}+\beta_{j+1}q_{j+1}e_{j}^{j\top},\label{eq:AQ}
  \\ 
  \mathcal{A}^{\top}P^{j} & = &
  P^{j}T^{j\top}+\gamma_{j+1}p_{j+1}e_{j}^{j\top},\label{eq:AP}\\ 
  P^{j\top}Q^{j} & = & I^{j},
  \label{eq:PQ}
\end{eqnarray} 
where $I^{j}$ indicates the $(j\times j)$ unit matrix, and $T^{j}$ is
the $(j\times j)$ tridiagonal matrix:
\begin{equation}
  T^{j}=
  \left(
    \begin{array}{ccccc}
      \alpha_{1} & \gamma_{2} & 0 & \cdots & 0\\
      \beta_{2} & \alpha_{2} & \gamma_{3} & 0 & \vdots\\
      0 & \beta_{3} & \alpha_{3} & \ddots & 0\\
      \vdots & 0 & \ddots & \ddots & \gamma_{j}\\
      0 & \cdots & 0 & \beta_{j} &
      \alpha_{j}
    \end{array}
  \right).
  \label{eq:tridiagonal}
\end{equation}
In the present case, because of the special block structure of the
Liouvillian super-operator and of the right-hand side appearing in
Eq. \eqref{eq:sup-1}, at each step of the Lanczos recursion one has
that $\mathcal{L}q_{j}$ is always orthogonal to $p_{j}$, so that,
according to Eq. \eqref{eq:lanczos-alpha}, $\alpha_{j}=0$. Let us now
rewrite Eq. (\ref{eq:AQ}) as:
\begin{equation}
  (\omega-\mathcal{A} )Q^{j}=Q^{j}(\omega-T^{j})-\beta_{j+1}q_{j+1}e_{j}^{j\top}.
  \label{eq:omega-A}
\end{equation}
By multiplying Eq. (\ref{eq:omega-A}) by $u^{\top}(\omega-\mathcal{A} )^{-1}$ on
the left and by $(\omega-T^{j})^{-1}e_{1}^{j}$ on the right, we
obtain: 
\begin{multline}
  u^{\top}Q^{j}(\omega-T^{j})^{-1}e_{1}^{j}= 
  u^{\top}(\omega-\mathcal{A} )^{-1}Q^{j}e_{1}^{j}-\\
  \beta_{j+1}u^{\top}(\omega-\mathcal{A})^{-1}q_{j+1}e_{j}^{j\top}
  (\omega-T^{j})^{-1}e_{1}^{j}.
  \label{eq:omega-A1}
\end{multline}
Taking the relation $Q_{j}e_{1}^{j}=q_{1}\doteq v$ into account,
Eq. (\ref{eq:omega-A1}) can be cast as: 
\begin{equation}
  g(\omega)=\left<\zeta^{j}|(\omega-T^{j})^{-1}e_{1}^{j}\right\rangle
  +\varepsilon_{j}(\omega),\label{eq:omega-A2}
\end{equation}
where: 
\begin{equation}
  \zeta^{j}=Q^{j\top}u\label{eq:zj-def}
\end{equation} 
is an array of dimension $j$, and:
\begin{equation}
  \varepsilon_{j}(\omega)=\beta_{j+1}\left\langle
    u|(\omega-\mathcal{A} )^{-1}q_{j+1}\right\rangle \left\langle
    e_{j}^{j}|(\omega-T^{j})^{-1}e_{1}^{j}\right\rangle
  .\label{eq:epsilon_j}
\end{equation}
is the error made when truncating the Lanczos chain at the $j$-th
step. Neglecting $\varepsilon_{j}(\omega)$ we arrive at the following
approximation to $g(\omega)$ defined in Eq. \eqref{eq:resolvent-def}
\begin{equation}
  \bar{g}_{j}(\omega) \approx \left\langle \zeta^{j}|(\omega-T^{j})^{-1}e_{1}^{j}
  \right\rangle \ .
  \label{eq:omega-App}
\end{equation} 
This approximation is the scalar product of two arrays of dimension
$j$: $\bar{g}_{j}(\omega)=\langle \zeta^{j}|\eta^{j}\rangle$, where $\eta^{j}$
is obtained by solving a tridiagonal linear system: 
\begin{equation}
  (\omega-T^{j})\eta^{j}=e_{1}^{j},
  \label{eq:tridiag-system}
\end{equation}
$T^{j}$ is the tridiagonal matrix of Eq. \eqref{eq:tridiagonal}, and
$\zeta^{j}$ is given by Eq. \eqref{eq:zj-def}. 

Three important practical observations should be made at this
point. The first is that solving tridiagonal systems is extremely
inexpensive (its operation count scales linearly with the system
size). The second is that the calculation of the sequence of vectors
$\zeta_{j}$ from Eq. \eqref{eq:zj-def} does not require the storage of
the $Q_{j}$ matrix. In fact, each component $\zeta^{j}$ is the scalar
product between one fixed vector ($u$) and the Lanczos recursion
vector $q^{j}$, and it can be therefore calculated on the fly along
the Lanczos recursion chain. We note that a slightly better approach
to evaluating Eq. \eqref{eq:omega-App} would be via the LU
factorization of the matrix $\omega-T^{j}$. If
$\omega-T_{j}=L_{\omega,j}U_{\omega,j}$, then
$\bar{g}(\omega)=\left\langle U_{\omega,j}^{-\top}\zeta^{j}|
  L_{\omega,j}^{-1}e_{1} \right\rangle $, 
%% \SB{Introduced braket
%%   notation: right Yousef?} 
%%  Yes -- correct.  added index j for L and U 
which can be implemented as the scalar
product of two sequences of vectors. We finally observe that the
components of $\zeta^j$ decrease rather rapidly as functions of the
iteration count, so that only a relatively small number of components
have to be explicitly calculated. This will turn out to be essential
for extrapolating the Lanczos recursion, as proposed and discussed in
Sec. \ref{sec:benchmark-enhance}. The components of
$\eta^{j}=(\omega-T^{j})^{-1}e^{j}$ also tend to decrease, although not
as rapidly. In fact this is used to measure convergence of the
Lanczos, or Arnoldi algorithms for solving linear systems, see, e.g.,
\cite{saad-book-lanczos}.

From the algorithmic point of view, much attention is usually paid in
the literature to finding suitable preconditioning strategies that
would allow one to reduce the number of steps that are needed to
achieve a given accuracy within a given iterative method
\citep{Olsen:1988lr}.  Although preconditioning can certainly help
reduce the number of iterations, it will in general destroy the nice
structure of the Lanczos factorization, Eq. \eqref{eq:AQ}, which is
essential to avoid repeating the time-consuming factorization
of the Liouvillian for different frequencies. In the next section we
will show how a suitable extrapolation of the Lanczos coefficient allows
for a substantial reduction of the number of iterations without
affecting (but rather exploiting) the nice structure of the Lanczos
factorization, Eqs. \eqref{eq:AP} and \eqref{eq:AQ}.

We conclude that the non-symmetric Lanczos algorithm allows one to
easily calculate a systematic approximation to the off-diagonal matrix
elements of the resolvent of a non-Hermitean matrix. It is easily
seen that, in the case of a diagonal matrix element, this same algorithm
would lead to a continued-fraction representation of the matrix element.
Although the representation of Eq. \eqref{eq:omega-A2}, which is
needed in the case of a non-diagonal element, is less elegant than
the continued-fraction one, its actual implementation is in practice
no more time consuming from the numerical point of view.

The idea of using the Lanczos algorithm to compute functions such the
one in Eq. \eqref{eq:resolvent-def} is not new. In control theory,
this function is called a \emph{transfer function} and it is used to
analyze the \emph{frequency response} of a system much like it is done
here.  Using the Lanczos algorithm for computing transfer functions
has been considered in, e.g., \cite{Saad-Contr,FeldmannFreund}. The
Lanczos and Arnoldi methods are also important tools in the closely
related area of model reduction in control theory, see, e.g.,
\cite{GrimmeAl96}.

\section{
  Benchmarking the new algorithm and enhancing its numerical
  performance
  \label{sec:benchmark-enhance}
} 

In this section we proceed to a numerical benchmark of the new
methodology against the test case of the benzene molecule, a system
for which several TDDFT studies already exist and whose optical
spectrum is known to be accurately described by ADFT
\citep{benzene-Yabana,Marques:2003,SB-Walker:06,benzene-Qian}. A
careful inspection of the convergence of the calculated spectrum with
respect to the length of the Lanczos chain allows us to formulate a
simple extrapolation scheme that dramatically enhances the numerical
performance of our method. All the calculations reported in the
present paper have been performed using the Quantum ESPRESSO
distribution of codes for PW DFT calculations \cite{QE}. Ground-state
calculations have been performed with the {\tt PWscf} code contained
therein, whereas TDDFT linear-response calculations have been
performed with a newly developed code, soon to be included in the
distribution. 

\subsection{Numerical benchmark}

The benchmark has been performed using the Perdew-Burke-Ernzerhof
(PBE) \citep{PBE} XC functional and USPP's
\citep{USPP-VdB:90,ralph-uspp,pp-quote} with a PW basis set up to a
kinetic energy cut-off of 30 Ry (180 Ry for the charge density). This
corresponds to a wavefunction basis set of about 25000 PW's, resulting
in a Liouvillian superoperator whose dimension is of the order of
750,000. Periodic boundary conditions have been used, with the
molecule placed horizontally flat in a tetragonal supercell of
$30\times30\times20$ $\mathrm{a}_0^{3}$. The absorption spectrum is
calculated as
$I(\omega)\propto\omega\mathrm{Im\left(\bar{\alpha}(\omega)\right)}$,
where $\bar{\alpha}$ is the spherical average (average of the diagonal
elements) of the molecular dipole polarizability. A small imaginary
part has been added to the frequency argument,
$\omega\rightarrow\omega+i\epsilon$, so as to regularize the
spectrum. This shift into the complex frequency plane has the effect
of introducing a spurious width into the discrete spectral lines. In
the continuous part of the spectrum, truncation of the Lanczos chain
to any finite order results in the discretization of the spectrum,
which appears then as the superposition of discrete peaks. The finite
width of the spectral lines has in this case the effect of broadening
spectral features finer than the imaginary part of the frequency, thus
re-establishing the continuous character of the spectrum. The optimal
value of the imaginary part of the frequency is slightly larger than
the minimum separation between pseudo-discrete peaks and depends in
principle on the details of the system being studied, as well as on
the length of the Lanczos chain and on the spectral region. Throughout
our benchmark we have rather arbitrarily set
$\epsilon=0.02\,\mathrm{Ry}$. Later in this section, we will see that
the length of the Lanczos chain can be effectively and inexpensively
increased up to any arbitrarily large size. By doing so, the distance
between neighboring (pseudo-) discrete states in the continuum
correspondingly decreases, thus making the choice of $\epsilon$
noncritical.

\begin{figure}
  \includegraphics[width=0.8\columnwidth]{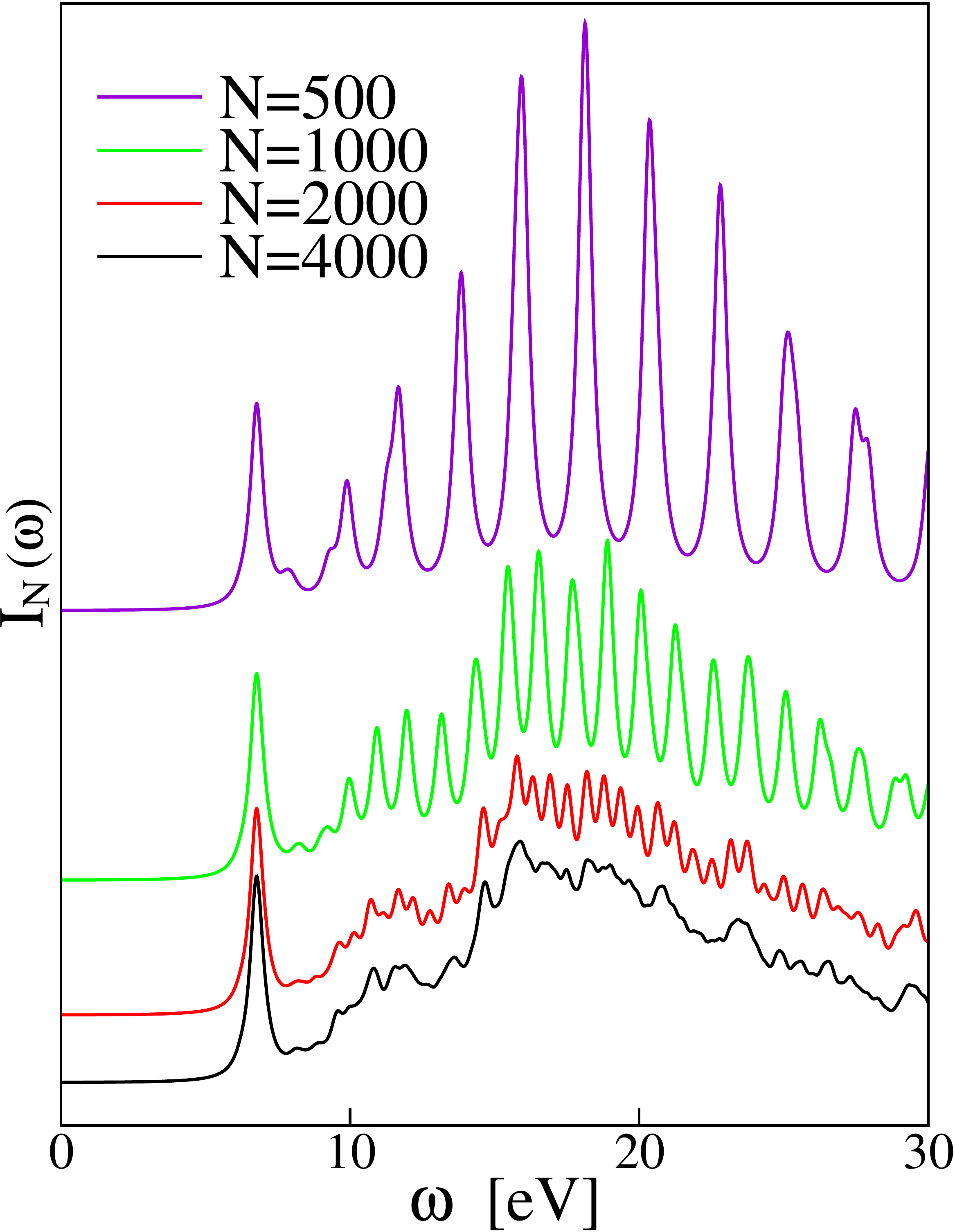}
  \caption{
    Absorption spectrum calculated using Lanczos method with
    ultrasoft pseudo-potentials. The figure shows the curve at
    different numbers of recursive steps; a vertical shift has been introduced
    for clarity. 
    }
  \label{fig:benzene_no_terminatore}
\end{figure}

In Fig. \ref{fig:benzene_no_terminatore} we report our results for the
absorption spectrum of the benzene molecule. The agreement is quite
good with both experimental data \citep{benzene-exp} and previous
theoretical work \citep{benzene-Yabana,Marques:2003,SB-Walker:06,benzene-Qian}.
\begin{figure}
  \includegraphics[width=0.8\columnwidth]{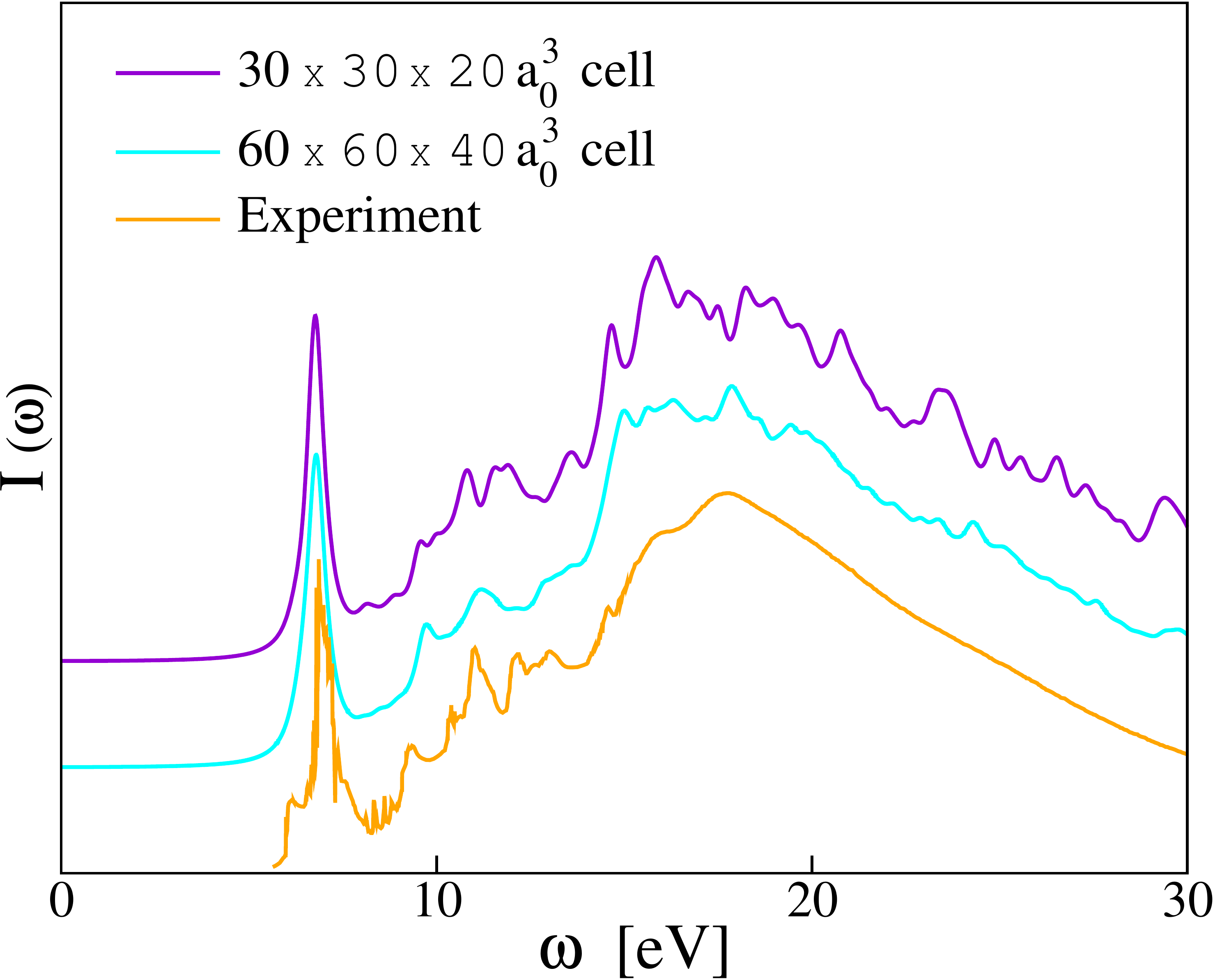}
  \caption{
    Comparison with experimental results of the converged spectrum of
    benzene for two different sizes of the cell; for the larger cell
    the structure in the continuum decreases and reproduces the
    experimental curve better.  Theoretical results have been scaled so as to
    obtain the same integrated intensity as experimental data. 
    \label{fig:large_small}
  }
\end{figure}
Above the ionization threshold the TDDFT spectrum displays a fine
structure (wiggles), which is not observed in experiments and that was
suggested in Ref. \citep{benzene-Yabana} to be due to size
effects associated to the the use of a finite simulation
cell. Finite-size effects on the fine structure of the continuous
portion of the spectrum are illustrated in Fig. \ref{fig:large_small}
where we display the spectrum of benzene as calculated using
two simulation cells of different size.

Our purpose here is not to analyze the features of the benzene
absorption spectrum, which are already rather well understood (see,
\emph{e.g.,} Ref. \citealp{SB-Walker:06}), nor to dwell on the
comparison between theory and experiment, but rather to understand
what determines the convergence properties of the new method and how
they can be possibly improved. The number of iterations necessary to
achieve perfect convergence lies in this case in-between 2000 and
3000: the improvement with respect to Ref. \citep{SB-Walker:06} is due
to the smaller basis set, made possible by the use of ultrasoft
pseudo-potentials, as discussed in Ref. \citep{ralph-uspp}. It is
worth noting that the convergence is faster in the low-energy portion
of the spectrum. This does not come as a surprise because the lowest
eigenvalues of the tridiagonal matrix generated by the Lanczos
recursion converge to the corresponding lowest eigenvalues of the
Liouvillian, and the lower the state the faster the convergence. 

A comparison between the performance of the new method with a more
conventional approach based on the diagonalization of the Liouvillian
is not quite possible because the two methodologies basically address
different aspects of a same problem. While the former addresses the
global spectrum of a specific response function, the latter focuses on
individual excited states, from which many different response
functions can be obtained, at the price of calculating all of the
individual excited states in a given energy range. It suffices to say
that it would be impractical to obtain a spectrum over such a wide
energy range as in Fig. \ref{fig:benzene_no_terminatore} by
calculating all the eigenvalues of a Liouvillian. Using a localized
basis set, which is the common choice in most implementations of
Casida's equations, it would be extremely difficult to resolve the
high lying portion of the one-electron spectrum with the needed
accuracy; using PW or real-space grid basis sets, instead, the
calculation of very many individual eigen-pairs of the Liouvillian
matrix whose dimension easily exceeds several hundreds thousands would
be a formidable task.

The comparison with time-propagation schemes is instead
straightforward and more meaningful. Typical time steps and total
simulation lengths in a time propagation approach are of the order of
$10^{-18}$ s, and $10^{-14}$ s, respectively, which amounts to about
10,000 time propagation steps \cite{ralph-uspp}. The computational
workload at each time step depends on the propagation algorithm. One
commonly used technique relies on a fourth-order Taylor expansion of the
propagator, together with so-called {\em enforced time-reversal
 symmetry} \cite{Castro-2004}. In this case, each time step requires
eight applications of the Hamiltonian to the KS orbitals and one
evaluation of the Hartree plus exchange-correlation potentials.
In the Lanczos approach, each step requires two applications of the
Hamiltonian and one evaluation of the potentials. Furthermore, the
response orbitals must be kept orthogonal to the ground-state
orbitals. This results in a computational effort which is sensibly
lower for one recursion step than for one time propagation
step. Considering both the larger number of propagation steps and the
more expensive workload at each step, we can conclude that our
approach is definitely more efficient than the time-propagation method
to compute linear response spectra.

\begin{figure}
  \includegraphics[width=0.9\columnwidth]{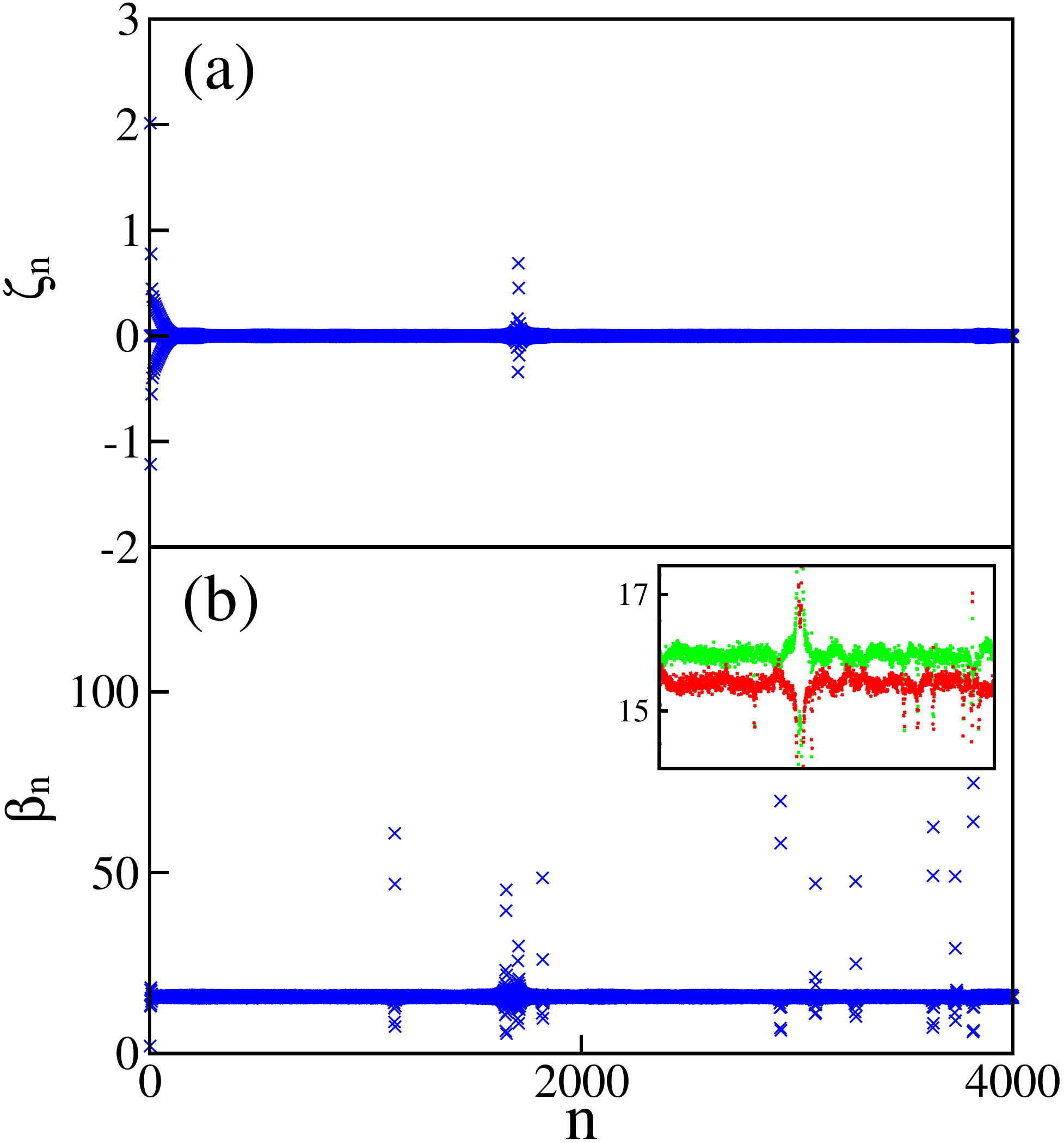}
  \caption{ 
    (a) Numerical behavior of the components of the $\zeta^{j}$
    vector given by equation \ref{eq:zj-def}.  Apart for some out of
    scale oscillation they tend rapidly to a value near zero. (b)
    Numerical behavior of $\beta_{j}$ coefficients given by
    Eq. (\ref{eq:betaj+1}). They tend rapidly to a constant value even
    if some larger scale oscillation is present. In the inset the same
    data are shown on a different scale and with different colors for
    odd (green) and even (red) coefficients.
  }
  \label{fig:beta_zeta}
\end{figure}

\subsection{
  Analysis
}

In Fig. \ref{fig:beta_zeta} we report the values of the $\beta$
coefficients and of the last component of the $\zeta$ vectors (see
Eqs. \ref{eq:betaj+1} and \ref{eq:zj-def}), as functions of the
Lanczos iteration count, calculated when the direction of both the
perturbing electric field and the observed molecular dipole are
parallel to each other and lying in the molecular plane (this would
correspond to calculating, say, the $xx$ component of the
polarizability tensor). It is seen that the $\zeta$ components rapidly
tend toward zero, whereas the $\beta$'s tend to a constant. Closer
inspection of the behavior of the latter actually shows that the
values of the $\beta$'s are scattered around two close, but distinct,
values for even and odd iteration counts. The $\gamma$ coefficients
(see Eq. \ref{eq:gammaj+1}) are in general equal to the $\beta$'s, and
only in correspondence with few iterative steps they assume a negative
sign.

All the calculated quantities, $\beta$, $\gamma$, and $\zeta$, are
subject to occasional oscillations off their asymptotic values. The
observed oscillations in the coefficients $\gamma_{j}$ and $\beta_{j}$
can be partly explained from their definitions, namely
Eqs. (\ref{eq:betaj+1}-\ref{eq:gammaj+1}). Note at first that there is
a risk of a division by zero in Eq. \eqref{eq:betaj+1}. The occurrence
of a zero scalar product $\langle\bar{q}|\bar{p}\rangle$ is known as a
\emph{breakdown}. Several situations can take place. A \emph{lucky}
breakdown occurs when one of the vectors $\bar{q}$ or $\bar{p}$ is
zero. Then the eigenvalues of the tridiagonal matrix are exact
eigenvalues of the matrix $\mathcal{A}$, as the space spanned by
$Q^{j}$ (when $\bar{q}=0$) becomes invariant under $\mathcal{A}$, or
the space spanned by $P^{j}$ (when $\bar{p}=0$) becomes invariant
under $\mathcal{A}^{\top}$. Another known situation is when neither
$\bar{q}$ nor $\bar{p}$ are zero but their inner product is exactly
zero. This situation has been studied extensively in the literature:
see, {\em e.g.}, \cite{Parlett-Taylor-Liu,Freund-Noel-Martin,%
  Cullum-Kerner-Willoughby}. One of the main results is that when this
breakdown takes place at step $j$ say, then it is often still possible
to continue the algorithm by essentially bypassing step $j$ and
computing $q_{j+2},p_{j+2}$, or some $q_{j+l},p_{j+l}$ where $l>1$,
directly. Intermediate vectors are needed to replace the missing
$q_{j+1},...q_{j+l-1}$ and $p_{j+1},...p_{j+l-1}$, but these vectors
are no longer bi-orthogonal, resulting in the tridiagonal matrix being
spoiled by {\em bumps} in its upper part. The class of algorithms
devised to exploit this idea are called {\em look-ahead Lanczos}
algorithms (LALAs), a term first employed in
\cite{Parlett-Taylor-Liu}.  Finally an \emph{incurable breakdown}
occurs when no pair $q_{i+l},p_{j+l}$ with some $l\ge 1$ can be
constructed which has the desired orthogonality properties. Note that
this type of breakdown cannot occur in the Hermitian Lanczos
algorithm, because it is a manifestation of the existence of vectors
in the right subspace (linear span of $Q^{j}$) that are orthogonal to
all the vectors of the left subspace (linear span of $P^{j}$), which
is impossible when these spaces are the same ($Q^{j}=P^{j}$ in the
Hermitian case). Clearly, exact breakdowns (inner product
$\langle\bar{q}|\bar{p}\rangle$ exactly equal to zero) almost never
occur in practice. Near breakdowns correspond to small values of these
inner products that determine the observed jumps in the coefficients
$\beta_{j},\gamma_{j}$.The components of the $\zeta_{j}$'s can also
show jumps in their magnitude since the vectors $q^{j}$ will
occasionally display large variations in norm. In finite-precision
arithmetics the occurrence and precise location of (near-) breakdowns
would also depend on the numerical details of the
implementation. Nevertheless in our experience the Lanczos recursion
always converges to the same final spectrum whose calculation is
therefore robust.

In order to understand what determines this robustness, we note that
our algorithm amounts to implicitly solving a linear system by an
iterative procedure based on a Lanczos scheme. This procedure is
mathematically equivalent to the Bi-Conjugate Gradient algorithm
(BiCG) \cite{saad-book-lanczos}. The observed robustness is therefore
consistent with what is known of BiCG \cite{saad-book-lanczos}. In
BiCG, the vector iterates lose their theoretical (bi-) orthogonality
and the scalars used to generate the recurrence may correspondingly
display very large oscillations, yet the solution of the linear
system, which is a linear combination of the vector iterates, usually
converges quite well. Because of this inherent robustness of the
algorithm, we preferred not to use any of the several available
LALAs. The shortcomings that these algorithms are designed to cure
not being critical, the marginal advantages that they may possibly
provide are outweighed by the drawback of losing the nice tridiagonal
structure of the $T_j$ matrices generated by them.

\begin{figure}
  \includegraphics[width=0.8\columnwidth]{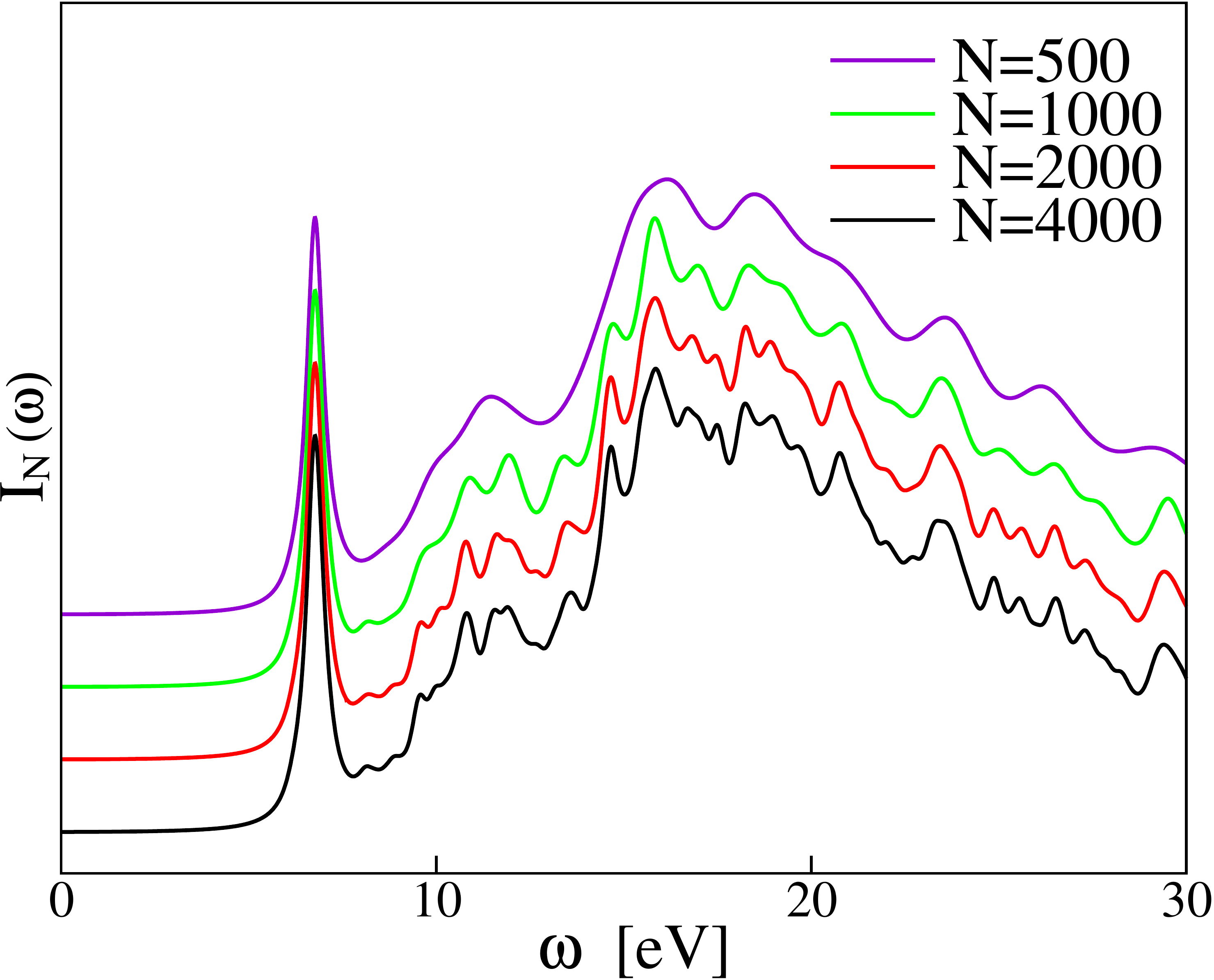} 
  \caption{
    Convergence of the absorption spectrum of benzene using the
    extrapolation procedure described in the text. After $N$
    iterations the components of $\zeta^{j}$ are set to zero and the
    $\beta$'s are extrapolated. The curves have been shifted vertically for
    clarity.
    \label{fig:benzene_terminatore} 
  }
\end{figure}
  
Another difficulty with generic Lanczos algorithms is the loss of
bi-orthogonality of the Lanczos vectors. As was mentioned earlier, in
exact arithmetic, the left and right Lanczos vectors are orthogonal to
each other. In the presence of round-off, a severe loss of
orthogonality eventually takes place.  This loss of orthogonality is
responsible for the appearance of so-called {\em ghost} or {\em
  spurious} eigen-pairs of the matrix to be inverted. As soon as the
linear span of the Lanczos iterates is large enough as to contain a
representation of an eigenvector to within numerical accuracy, the
subsequent steps of the Lanczos process will tend to generate replicas
of this eigenvector.  At this point the Lanczos bases (left or right
spaces) become linear dependent to within machine precision. From the
point of view of solving the systems $ (\omega - \mathcal{A}) x = v $,
the effect of these replicated eigenvalues is not very
important. Indeed, when thinking in terms of the BiCG algorithm, after
the underlying sequence of approximations $x^j = Q^j (\omega -
T^j)^{-1} e_1^j$ obtained from the BiCG algorithm converges to $ x
= (\omega - \mathcal{A})^{-1} v $, further iterations will only add
very small components to $x^j$. As a result the contributions of these
replicas are bound to be negligible and this is observed in
practice. Thus, ghost eigenvectors have zero (or very small)
oscillator strengths and their contribution to the wanted inner products
$\langle u | x^j \rangle $, which approximate $g(\omega) $ in
Eq. \eqref{eq:resolvent-def}, will be negligible in general.
 
\subsection{Extrapolating the Lanczos recursion chain}

The fast decrease of the components of $\zeta^{j}$ implies that the
quality of the calculated spectrum depends only on the first few
hundreds of them. Specifically, if we set the components of the $\zeta^j$
vector equal to zero in Eq. \eqref{eq:omega-A2} after, say, 300-400
iterations, but we keep the dimension of the tridiagonal matrix,
$T^{j}$, of the order of 2-3000, the resulting spectrum appears to be
still perfectly converged. Unfortunately, a relatively large number of
iterations seems to be necessary to generate a tridiagonal matrix of
adequate dimension. The regular behavior of the $\beta$'s for large
iteration counts suggests an inexpensive strategy to extrapolate the
Lanczos recursion. Let us fix the dimension of the tridiagonal matrix
in Eq. \eqref{eq:omega-A2} to some very large value (say, $N^\star=10000$),
and define an effective $\zeta^{j}$ vector, $\zeta_{N}^{N^\star}$, and
$T^{j}$ matrix, $T_{N}^{N^\star}$, by setting the $k$-th component of
$\zeta_{N}^{N^\star}$ equal to zero for $k>N$, and the $k$-th component of
$\beta$ equal to the appropriate estimate of the asymptotic value for
odd or even iteration counts, obtained from iterations up to $N$.  In
general, as previously noted, it very seldom occurs that 
$\gamma_{j}$ and $\beta_{j}$ have a different sign, and we found that
that extrapolating them to the same positive value does not
invalidate significantly the accuracy of the extrapolation.

In Fig. \ref{fig:benzene_terminatore} we display the spectra,
$I_{N}(\omega)$, obtained from the extrapolation procedure just
outlined, which from now on will be referred to as the {\em bi-constant
  extrapolation} of the Lanczos coefficients. One sees that the
extrapolated spectrum is at perfect convergence already for a very
modest value of $N$ in between $N=500$ and $N=1000$, a substantial
improvement with respect to the results shown in
Fig. \ref{fig:benzene_no_terminatore}.  Note that this extrapolation
procedure, although necessarily approximate, offers a practical
solution to the problem of recovering a continuous spectrum from a
limited number of recursion steps. As the dimension of the tridiagonal
matrix appearing in Eq. \eqref{eq:omega-A2} can be made arbitrarily
large at a very small cost, the distance between neighboring
pseudo-discrete eigenvalues in the continuous part of the spectrum can
be made correspondingly small, thus allowing to chose the imaginary
time of the frequency basically as small as wanted.

A qualitative insight into the asymptotic behavior of the Lanczos
recursion coefficients can be obtained from the analogy with the
continued-fraction expansion of the local density of states (LDOS) for
tight-binding (TB) Hamiltonians, a problem that has been the breeding
ground for the application of Lanczos recursion methods to
electronic-structure theory
\citep{haydock-72-1,haydock-75,recursion-ssp,pastori-recursion}. Since
the late seventies it has been known that the coefficients of the
continued-fraction expansion of a \emph{connected} LDOS asymptotically
tend to a constant---which equals one fourth of the band
width---whereas they oscillate between two values in the presence of a
gap: in the latter case the average of the two limits equals one
fourth of the total band width, whereas their difference equals one
half the energy gap \citep{turchi:JPC-1982}.  These results can be
easily verified in the case of a 1D TB Hamiltonian with constant
hopping parameter, $\beta$, which leads to the continued fraction:
\begin{eqnarray}
  g(\omega) & = & \frac{{\displaystyle 1}}{{\displaystyle 
      \omega-\frac{{\displaystyle \beta^{2}}}{{\displaystyle
          \omega-\frac{{\displaystyle \beta^{2}}}{{\displaystyle
              \omega-}\cdots}}}}} \nonumber \\ 
  & = &
  \frac{\omega\pm\sqrt{\omega^{2}-4\beta^{2}}}{2\beta^{2}},
\end{eqnarray}
where the sign has to be chosen so as to make the Green's function
have the proper imaginary part. In this case, one sees that the
imaginary part of the Green's function (which equals the LDOS) is
non-vanishing over a band that extends between $-2\beta$ and
$2\beta$. In the case were consecutive hopping parameters of the
recursion chain oscillate between two values, $\beta_{1}$ and
$\beta_{2}$, which we assume to be positive, the resulting Green's
function reads:
\begin{align}
  g(\omega) & = \frac{{\displaystyle 1}}{{\displaystyle
      \omega-\frac{{\displaystyle \beta_{1}^{2}}}{{\displaystyle
          \omega-\frac{{\displaystyle \beta_{2}^{2}}}{{\displaystyle
              \omega-\cdots}}}}}} \nonumber \\ 
  & =
  \frac{\omega^{2}+\beta_{1}^{2}-\beta_{2}^{2}\pm\sqrt{(\omega^{2}+
      \beta_{1}^{2}-\beta_{2}^{2})^{2}-4\omega^{2}
      \beta_{1}^{2}}}{2\omega\beta_{1}^{2}}.
\end{align} 
in this case we obtain two bands between $|\beta_{1}-\beta_{2}|$ and
$\beta_{1}+\beta_{2}$ and between $-(\beta_{1}+\beta_{2})$ and
$-|\beta_{1}-\beta_{2}|$.

\begin{figure}
  \includegraphics[width=1.0\columnwidth]{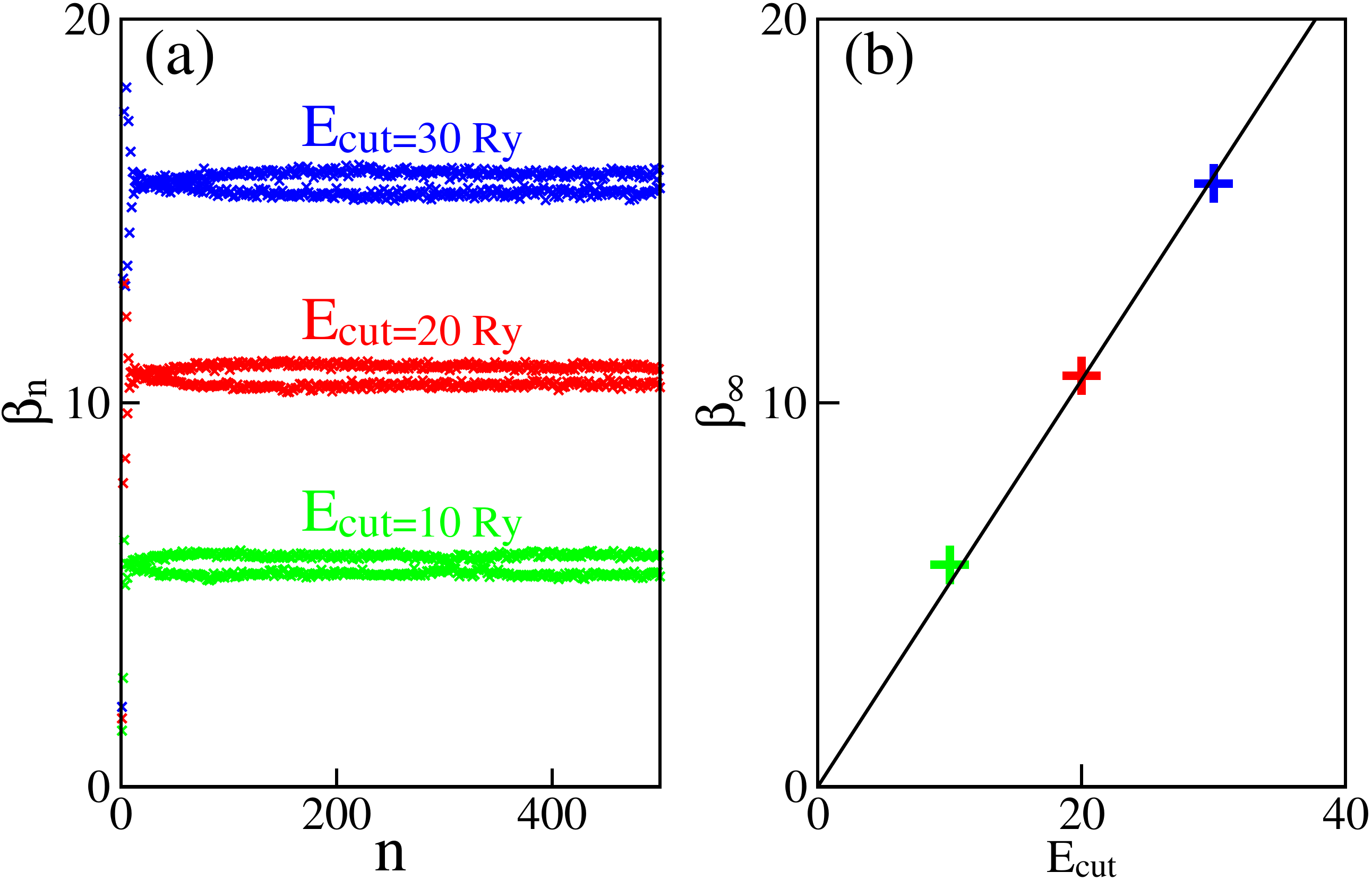} 
  \caption{ (a) Behavior of $\beta$'s coefficients of benzene for different
    values of the kinetic energy cut-off. (b) The asymptotic values
    $\beta_{\infty}$ plotted as a function of the kinetic energy
    cut-off; the figure shows that they can be connected by a straight
    line with slope of about $0.5$.
    \label{fig:beta} 
  }
\end{figure}

In our case, the relevant band width of the Liouvillian super-operator
extends from minus to plus the maximum excitation energy. In a PP-PW
pseudo-potential scheme, in turn, the latter is of the order of the
PW kinetic-energy cutoff, $E_{cut}$, whereas the gap is of the order
of twice the optical gap, $\Delta$. We conclude that the asymptotic
values for the $\beta$ and $\gamma$ coefficient of the Liovillian
Lanczos chain are: $\frac{\beta_{1}+\beta_{2}}{2}\approx\frac{E_{cut}}{2}$
and $|\beta_{1}-\beta_{2}|\approx\Delta$. In Fig. \ref{fig:beta}a we
report the behavior of the values of the $\beta$ coefficients of
the Liouville Lanczos chain calculated for benzene, \emph{vs.} the
iteration count, for different plane-wave kinetic-energy cutoffs. In
Fig. \ref{fig:beta}b the average asymptotic value is plotted against
the kinetic-energy cutoff, demonstrating a linear dependence
$\beta_\infty \approx\frac{1}{2} E_{cut}$, in remarkable agreement
with the qualitative analysis described above. Also the difference
between the asymptotic values for odd and even iteration counts
($|\beta_\infty^{odd} - \beta_\infty^{even}|\approx 0.46~\textrm{Ry}$)
is in remarkable qualitative agreement with the optical gap
($\Delta=0.38~\textrm{Ry}$).

\section{ 
  Application to large molecules: fullerene and chlorophyll a 
  \label{sec:applications} 
}

In order to demonstrate the applicability of our methodology to large
molecular systems, we present now the results obtained for the
prototypical cases of fullerene C$_{60}$ and chlorophyll a.

\begin{figure}[h]
  \includegraphics[width=0.8\columnwidth]{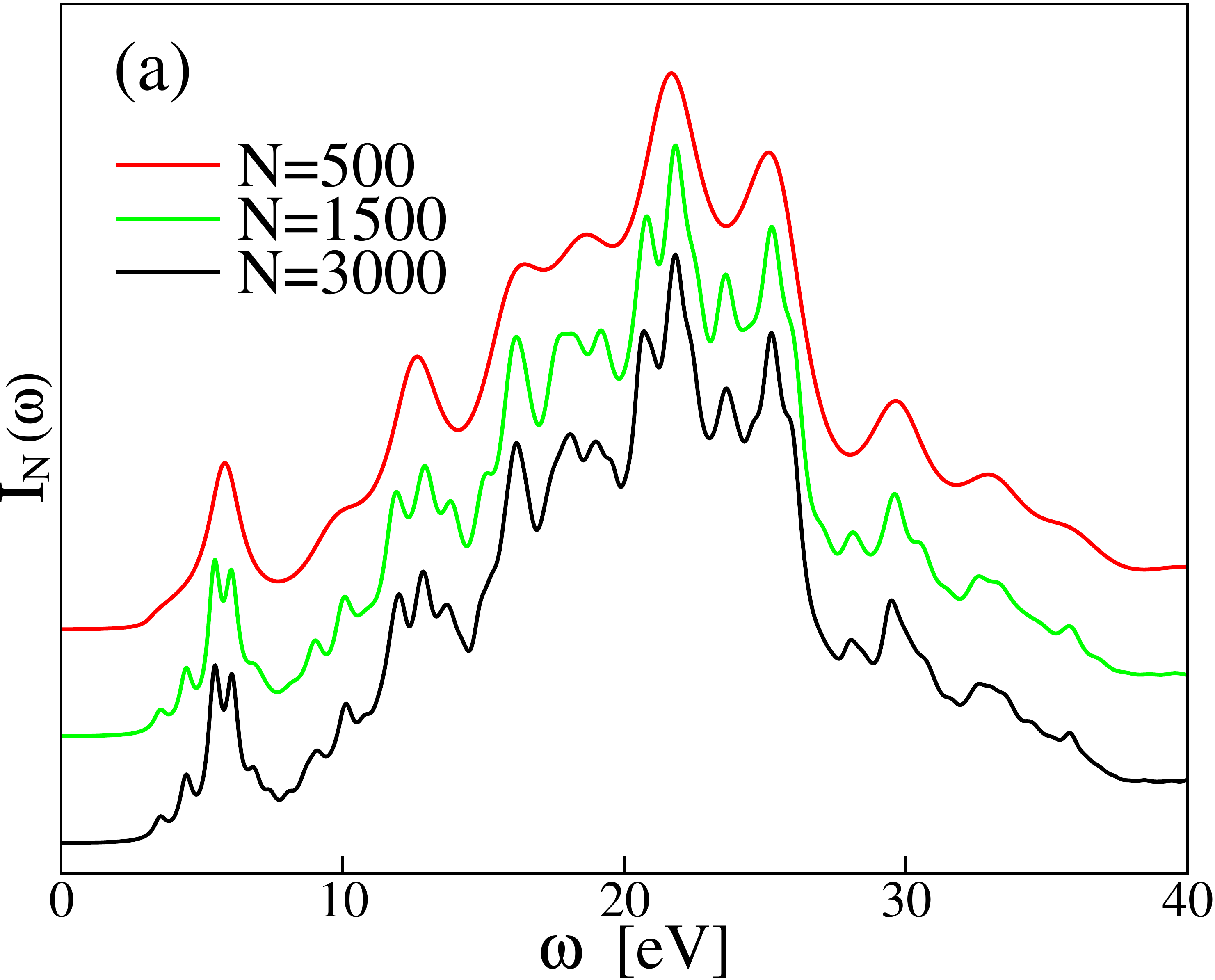}
  \\
  \includegraphics[width=0.8\columnwidth]{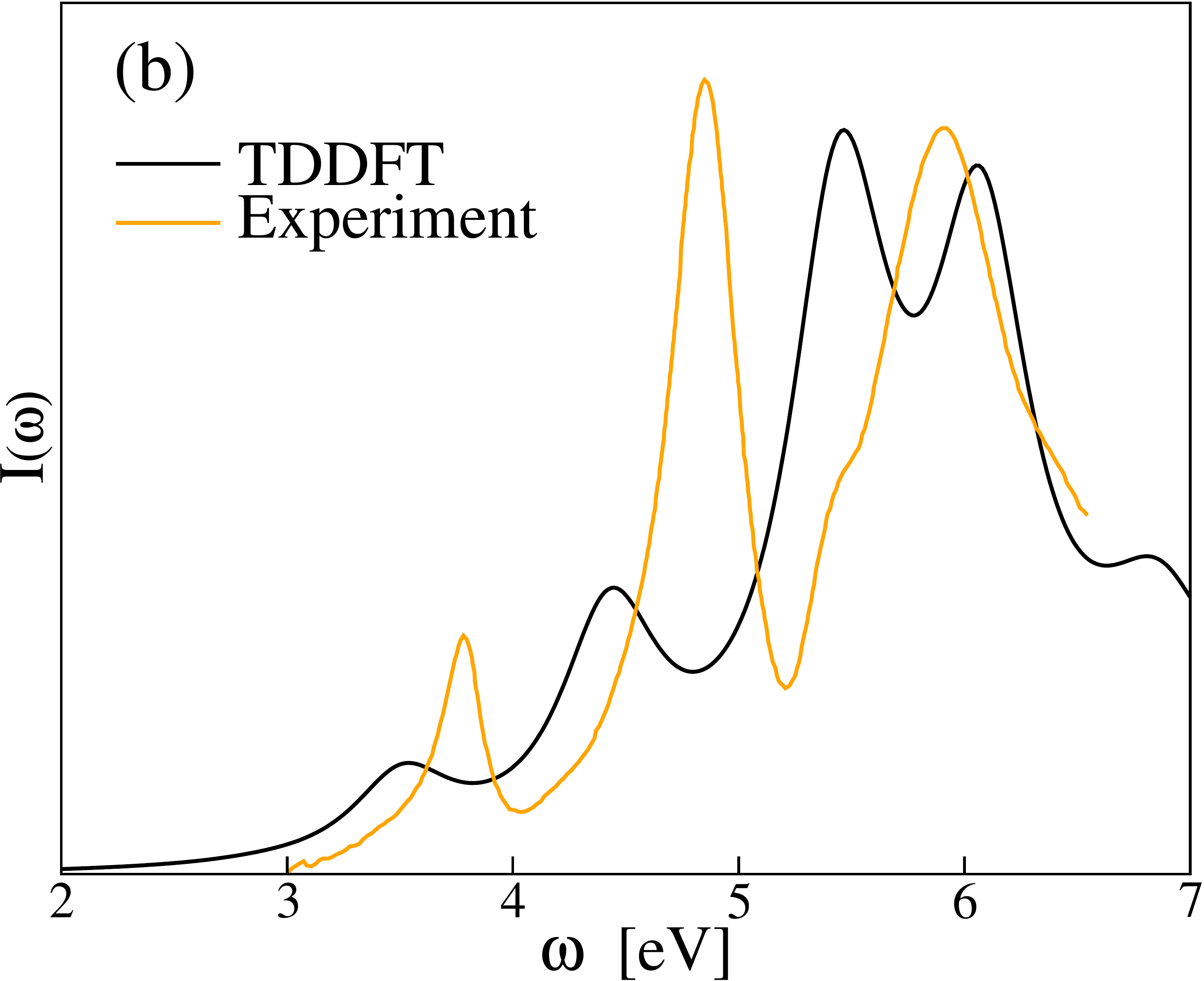}
  \caption{ (a) Convergence of the absorption spectrum of fullerene
    calculated between 0 and 40 eV. The curves have been shifted vertically for
    clarity. (b) The fully converged absorption
    spectrum of fullerene compared with experimental results
    \citep{fullerene-th-exp} in the energy range between 2 and 7
    eV. For comparison purposes TDDFT results have been rescaled
    in order to have the first transition peak at the same height as
    that of experimental results. Theoretical results 
    have been scaled so as to obtain the same integrated intensity
    as experimental data. \label{fig:fullerene}
  }
\end{figure}

Let us begin with fullerene, a molecule whose spectrum has already
been the subject of extensive experimental
\citep{fullerene-th-exp,fullerene-exp} and theoretical
\citep{elstner-fullerene,martin-fullerene,fullerene-th-exp,%
  Yabana-Bertsch:96,benzene-Yabana,ordejon-fullerene} studies. Our
calculations have been performed with the molecule lying in a cubic
super-cell with side length of 35 $\mathrm{a}_0$, using the PBE XC functional.
Ultra-soft pseudo-potentials \cite{pp-quote} have been used, with a PW
basis set with a kinetic energy cut-off of 30 Ry for the wavefunctions
and 180 Ry for the charge density. This corresponds to almost 60,000
PW's, with a dimension of the full Liouvillian exceeding 14
millions. The Lanczos recursion is explicitly computed up to
different orders, $N$, as indicated in Sec. \ref{sec:Lanczos}, and
then extrapolated up to $N^\star=20000$, as discussed in
Sec. \ref{sec:benchmark-enhance} (this value has been chosen rather
arbitrarily because both the numerical workload and the resulting
accuracy depends very little on it, as long as it is large enough). In
order to regularize the solution of the tridiagonal linear system,
Eq. \eqref{eq:tridiag-system}, the spectrum has been calculated at
complex frequencies whose imaginary part is (also rather arbitrarily)
taken as $\epsilon=0.02\textrm{ Ry}$. In Fig. \ref{fig:fullerene}a we
report the calculated absorption spectrum between 0 and 40 eV. We see
that, upon bi-constant extrapolation, the calculated spectrum is
already very good after as few as 500 iterations, and practically
indistinguishable from convergence after 1500 iterations. The
resulting spectrum depends very little on the precise choice of
$\epsilon$ as long as its value is smaller than the distance
between neighboring eigenvalues of the tridiagonal matrix of
Eq. \eqref{eq:tridiag-system} (this distance goes to zero in the
continuous portion of the spectrum as $N^\star$ grows large), and
larger than the desired resolution of the calculated spectrum.

The overall shape of our calculated spectrum is in substantial
agreement with that calculated in Refs. \citep{martin-fullerene,
  benzene-Yabana,Yabana-Bertsch:96} using the real-time approach to
TDDFT. In spite of the small atomic basis set used in Ref.
\citep{martin-fullerene}, the number of integration steps that was
found to be necessary to reach an acceptable accuracy (6000) is rather
larger than ours. In Refs. \citep{benzene-Yabana,Yabana-Bertsch:96}
where a real-space grid representation of the KS equations was
adopted, instead, the number of time steps employed is one to two
orders of magnitude larger than ours (30-40,000). Considering that
several $H\psi$ products are necessary at each time step in real-time
approaches, whereas only two are needed at each Lanczos recursion, we
see that our combined use of the Liouville-Lanczos algorithm with
bi-constant extrapolation and ultra-soft pseudopotentials with plane
waves allows for a substantial reduction of the numerical workload,
while keeping the full accuracy allowed by the XC functionals
currently available.

The absorption spectrum of C$_{60}$ is characterized by a low-lying
and well structured portion (between, say, 3 and 7 eV) dominated by
$\pi\to\pi^\star$ transitions, followed by a broader feature between
14 eV and 27 eV determined by transitions from both $\sigma$ and $\pi$
molecular orbitals. In Fig. \ref{fig:fullerene}(b) we compare our
converged spectrum with the experimental results of
Ref. \citep{fullerene-th-exp}.  Despite a slight redshift compatible
with that found in the calculations of Ref. \citep{fullerene-th-exp},
the overall shape of the TDDFT spectrum is in good agreement with
experiment. Note that the theoretical results reported in
Ref. \citep{fullerene-th-exp}, which were obtained by calculating
individual eigen-pairs of the Casida's equation, could hardly be
extended to such a broad energy range as covered in the present
calculation, because too many lines would have to be calculated.

\begin{figure}[h]
  \includegraphics[width=0.8\columnwidth]{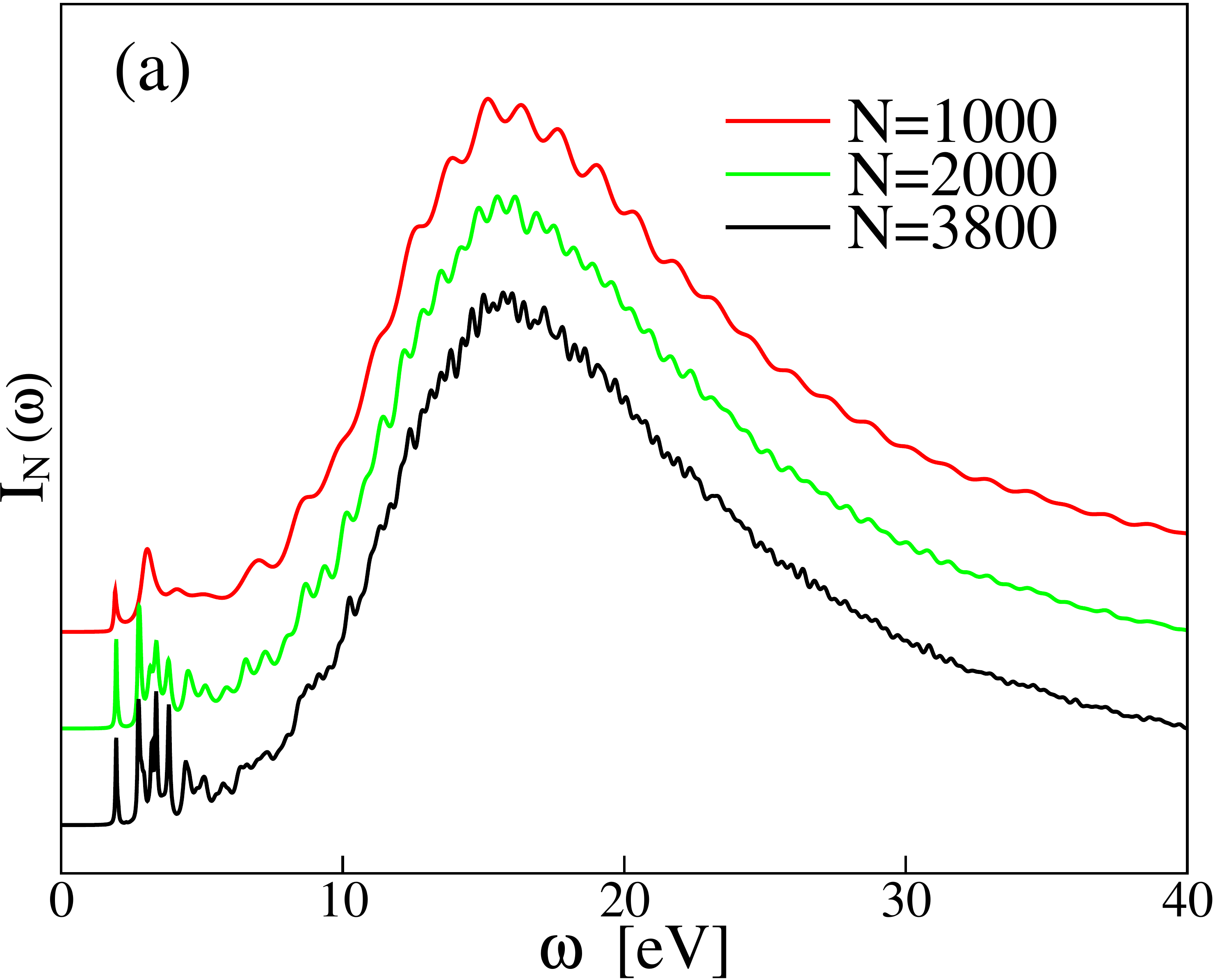}
  \includegraphics[width=0.8\columnwidth]{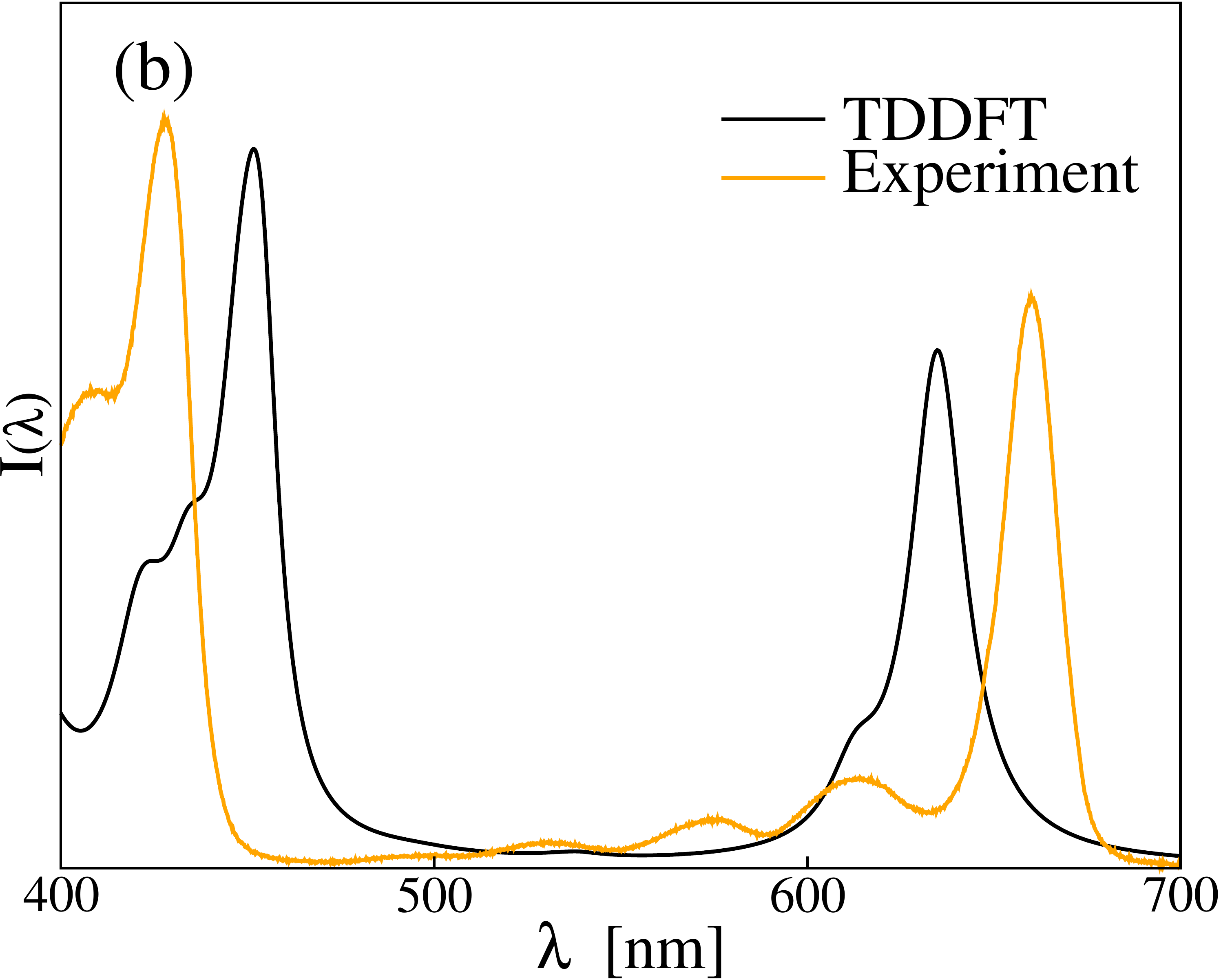}
  \caption{ (a) Convergence of the chlorophyll absorption spectrum
    between 0 and 40 eV. The curves have been shifted vertically for
    clarity. (b) Chlorophyll absorption spectrum in the
    visible region for wavelengths between 400 and 700 nm compared with the
    experimental data in di-ethyl ether of
    Ref.\citep{ch-exp}. Theoretical results  
    have been scaled so as to obtain the same integrated intensity
    as experimental data.
    \label{fig:clorofilla}
  }
\end{figure}

An even more challenging test is chlorophyll, a molecule which is of
fundamental importance for life on Earth since it is responsible for
the photosynthetic process. There are several different forms of this
molecule, and we will focus on chlorophyll a
$(\mathrm{C_{55}H_{72}MgN_{40}})$. Historically the interpretation of
the visible spectrum of chlorophyll relies on the 4-orbital Gouterman
model of porphyrins \citep{gouterman} in which only the two highest
occupied molecular orbitals and the two lowest unoccupied molecular
orbitals are considered. In the last few years there have been several
calculations of its low energy spectrum relying on different {\em ab
  initio} techniques
\citep{ch-sundholm,ch-sundholm-2,ch-linnanto,ch-dahlbom,ch-linnanto-2,ch-cai}.
Despite the fact that TDDFT seems to produce spurious charge transfer
states in the visible region \citep{ch-dahlbom}, according to our
calculations the overall shape of the low energy part of spectrum
seems to be correctly predicted. Our calculations have been performed
using a super-cell of dimensions $35\times45\times55~\mathrm{a}_0^{3}$
with the PW91 XC functional \citep{PW91} and USPP's
\cite{pp-quote}. Molecular orbitals where expanded in PW's up to a
kinetic energy cut-off of 30 Ry, while 180 Ry were used for the charge
density. The PW basis sets consists of more than 120000 PW's, while
the dimension of the Liovillian superoperator exceeds 42 millions. In
this case the imaginary part of the frequency was set to $\epsilon=
0.002~\mathrm{Ry}$ to better compare the results with experiments.  In
Fig. \ref{fig:clorofilla}(a) we display the convergence of the
spectrum with respect to the number of Lanczos steps, using the usual
bi-constant extrapolation of the coefficients, as calculated over a
wide range of energy between 0 and 40 eV.  In
Fig. \ref{fig:clorofilla}(b) we compare the visible part of the
spectrum calculated in this work with the experimental results
obtained in diethyl solution in Ref. \citep{ch-exp}. The agreement
with experiment is clearly good but the Soret (B) band located in the
indigo region of the spectrum at 430 nm is slightly red-shifted in the
calculation, while the red band (Q) has an opposite, blue-shifted
behavior. How much of this discrepancy has to be attributed to the
limitations of the AXCA alone, or to a combination of them with the
neglect of solvation effects remains to be ascertained.

\section{
  Conclusions
  \label{sec:conclusions}
} 

In this paper we have presented a new algorithmic approach to
linearized TDDFT that combines the advantages of the more conventional
real-time and Casida's eigenvalue methods, while avoiding many of
their drawbacks. This approach results from the combination of many
elements which are individually not new in different communities,
ranging from condensed matter and quantum chemistry, to control
theory/engineering and signal processing. In particular it is the
natural extension to the dynamical regime of density-functional
perturbation theory, a method made popular in the condensed-matter
community by the calculation of static properties (such as dielectric,
piezoelectric, elastic) and by the calculation of phonons and related
properties in crystals. The main features of the new method are that
it is tailored to the calculation of {\em specific responses} to {\em
  specific perturbations} and that the computational burden for the
calculation of the {\em complete} spectrum of a given response
function in a wide frequency range is comparable to that of a {\em
  single} static ground-state or response-function calculation. We
believe that, from the algorithmic point of view, the new method is
close to optimal in its application range and that it opens thus the
way to the simulation of the dynamical properties of large and very
large molecular and condensed-matter systems. Assuredly, it cannot
yield any better results than granted by the quality of the XC
functional used to implement it. Devising new XC functionals capable
of properly describing the electron-hole interaction responsible, {\em
  e.g.}, of Rydberg and excitonic effects in the low-lying portion of
the spectrum of molecular and extended systems, respectively, remains
a major problem to be addressed and solved.

\begin{acknowledgments}
  SB wishes to thank Renata Wentzcovitch for hospitality at the
  University of Minnesota Department of Chemical Engineering and
  Materials Science and vLab, where some of the work reported in this
  paper was started in Summer 2005. RG and SB thank Brent Walker and
  Marco Saitta for participating in the early developments of the ideas
  on which this paper is founded. DR gratefully acknowledges useful
  conversations with Giuseppe Pastori Parravicini and correspondence
  with Liana Martinelli, the former also bearing the
  responsibility for making SB aware of the power and beauty of Lanczos
  methods. 
\end{acknowledgments}

\bibliographystyle{apsrev}
\bibliography{turbo}

\end{document}